\documentclass[]{article}

\usepackage{times}
\usepackage{soul}
\usepackage{url}
\usepackage[hidelinks]{hyperref}
\usepackage[utf8]{inputenc}
\usepackage[small]{caption}
\usepackage{graphicx}
\usepackage{amsmath}
\usepackage{amsthm}
\usepackage{booktabs}
\usepackage[switch]{lineno}

\usepackage{amsfonts}                                             
\usepackage[ruled]{algorithm2e}                                   
\usepackage{color}                                                
\usepackage{colonequals}                                          
\usepackage{amssymb}
\usepackage{framed}                                               
\usepackage{enumerate}
\author{
Carlos Aguilera-Ventura$^1$,   
Xinghan Liu$^2$,
Emiliano Lorini$^1$ \\
\text{and}
Dmitry Rozplokhas$^2$\\
\normalsize{$^1$IRIT, CNRS, Toulouse University, France}\\
\normalsize{$^2$TU Wien, Austria} \\
}

\date{}

\newcommand*{\ArbProps}{Z}                                        
\newcommand*{\base}{C}                                            
\newcommand*{\ctx}{U}                                             
\newcommand*{\isdef}{\stackrel{\text{\tiny def}}{=}}              
\newcommand*{\lang}{\mathcal{L}}                                  
\newcommand*{\langP}{\lang_{\mathsf{PROP}}}                       
\newcommand*{\langequat}{\lang_{\mathsf{EQ}}}                       
\newcommand*{\langE}{\lang_0 }                         
\newcommand*{\langDI}{\lang }                      
\newcommand*{\langQBF}{\lang_{\textsc{QBF}}}                      
\newcommand*{\lequ}{\leftrightarrow}                              
\newcommand*{\limp}{\rightarrow}                                  
\newcommand*{\ltri}{\triangle}                                    
\newcommand*{\Nats}{\mathbb{N}}                                   
\newcommand*{\Terms}{\mathit{Term}}                                   
\newcommand*{\Interv}[1]{\mathit{Int}_{#1} }                                   
\newcommand*{\locint}[2]{ #1\leftrightarrow#2  }                                   
\newcommand*{\globint}{E  }                                   
                                  
\newcommand*{\conjfunct}[1]{ \widehat{#1}  }                                   %
\newcommand*{\prop}{p}                                            
\newcommand*{\term}{\lambda}                                            
\newcommand*{\propb}{q}                                           
\newcommand*{\Props}{\mathbb{P}}                                  
\newcommand*{\sat}{\models}                                       
\newcommand*{\state}{S}                                           
\newcommand*{\States}{\mathbf{S}}                                 
\newcommand*{\Models}{\mathbf{M}}                                 
\newcommand*{\subbase}{X}                                         
\newcommand{\tuple}[1]{({#1})}                                    
\newcommand*{\univCtx}{\mathbf{S}}                                
\newcommand*{\univCtxEq}{\mathbf{S}_{\mathit{Eq}}}                                
\newcommand*{\val}{V}                                             
\renewcommand{\phi}{\varphi}

\newcommand{\Omit}[1]{}                                           

\newcommand*{\engfunct}{\mathit{end}  }                                            
\newcommand*{\exgfunct}{\mathit{exo}  }                                            

\newcommand*{\butcause}{\mathsf{But}  }                                            

\newcommand{\causegraph}{G  }

\newcommand{\graphneigh}{\mathcal{P} }
  \newcommand{\nodeset}{\mathit{N} }

          \DeclareSymbolFont{symbolsC}{U}{txsyc}{m}{n}
\DeclareMathSymbol{\boxRight}{\mathrel}{symbolsC}{128} 
\DeclareMathSymbol{\diamondRight}{\mathrel}{symbolsC}{132}
\DeclareMathSymbol{\boxRightPoint}{\mathrel}{symbolsC}{130}
\DeclareMathSymbol{\diamondRightPoint}{\mathrel}{symbolsC}{134}

\newcommand{\condmodal}[2]{  #1 \boxRight  #2   }

\newcommand{\putaway}[1]{  }

\renewcommand{\phi}{\varphi}
\newcommand{\suchthat}{ ~ : ~ }

\renewcommand{\phi}{\varphi}

\newcommand{\relstateinterv}[1]{\Rightarrow^{#1}}

\newcommand{\forallqbf}[2]{\forall #1.\;#2}
\newcommand{\cloneVar}{\xi}

\newcommand{\subformulas}[1]{\textit{Sub}\mathcal{F}(#1)}
\renewcommand{\O}[1]{\mathcal{O}(#1)}

\newcommand{\univCtxRel}[1]{\univCtx^{#1}}
\newcommand{\inputform}{\psi}
\newcommand{\qbfindi}{i}
\newcommand{\qbfindii}{j}
\newcommand{\qbfindiii}{k}
\newcommand{\stateVars}[1]{X^{#1}}
\newcommand{\stateVarsi}{X^{\qbfindi}}
\newcommand{\stateVarsii}{X^{\qbfindii}}
\newcommand{\stateVarsiii}{X^{\qbfindiii}}
\newcommand{\statevarbase}[2]{b^{#1}_{#2}}
\newcommand{\statevarbasei}{\statevarbase{\qbfindi}{\omega}}
\newcommand{\statevarbaseii}{\statevarbase{\qbfindii}{\omega}}
\newcommand{\statevarbaseiii}{\statevarbase{\qbfindiii}{\omega}}
\newcommand{\statevarval}[2]{v^{#1}_{#2}}
\newcommand{\statevarvali}{\statevarval{\qbfindi}{\prop}}
\newcommand{\statevarvalii}{\statevarval{\qbfindii}{\prop}}
\newcommand{\statevarvaliii}{\statevarval{\qbfindiii}{\prop}}
\newcommand{\qbfVal}{Y}
\newcommand{\qbfVali}{Y^{\qbfindi}}
\newcommand{\qbfValii}{Y^{\qbfindii}}
\newcommand{\qbfValiii}{Y^{\qbfindiii}}

\newcommand{\qbfValToState}[2]{\mathcal{S}^{#1}(#2)}

\newcommand{\maprel}{\pi}
\newcommand{\modelsQBF}{\models_{\text{QBF\small}}}

\newcommand{\eqdef}{\stackrel{\mathit{def} }{=}}

\newcommand{\litVars}[1]{L^{#1}}
\newcommand{\litvarpos}[2]{l^{#1}_{#2}}
\newcommand{\litvarneg}[2]{l^{#1}_{\neg #2}}

\renewcommand{\O}[1]{\mathcal{O}(#1)}

\newcommand{\closest}{\mathtt{Closest}}
\newcommand{\qbfPredStyle}[1]{\mathsf{#1}}
\newcommand{\qbfSat}{\qbfPredStyle{Sat}}
\newcommand{\qbfInit}{\qbfPredStyle{Init}}

\newcommand{\qbfCloser}{\qbfPredStyle{Closer}}
\newcommand{\qbfClosereq}{\qbfPredStyle{Closereq}}
\newcommand{\qbfClosest}{\qbfPredStyle{Closest}}
\newcommand{\qbfState}{\qbfPredStyle{State}}
\newcommand{\qbfEq}{\qbfPredStyle{Eq}}

\newcommand{\qbfTerm}{\qbfPredStyle{Term}}
\newcommand{\qbfTermSat}{\qbfPredStyle{TermSat}}
\newcommand{\qbfTermExo}{\qbfPredStyle{TermExo}}
\newcommand{\qbfTermRev}{\qbfPredStyle{InvTerm}}
\newcommand{\qbfTermMerge}{\qbfPredStyle{MergeTerms}}
\newcommand{\qbfTermVarsSubset}{\qbfPredStyle{TermVars^\subset}}
\newcommand{\qbfTermVarsSubseteq}{\qbfPredStyle{TermVars^\subseteq}}
\newcommand{\qbfTermClosest}{\qbfPredStyle{TermClosest}}
\newcommand{\qbfTermCounterfactual}{\qbfPredStyle{TermCtrfact}}
\newcommand{\qbfActualCause}{\qbfPredStyle{ActualCause}}

\usepackage{tcolorbox}
\tcbuselibrary{skins, breakable}
\newtcolorbox{formal}{
	colback=red!5!white,
	colframe=red!75!black,
	breakable,
	enhanced jigsaw,
	pad at break* = 0mm,
	boxrule = 1mm,
	arc = 2mm,
	boxsep = 0mm,
	left* = 0mm,
	grow to left by = 3mm,
	right* = 0mm,
	grow to right by = 3mm
}
\newtheorem{theorem}{Theorem}

\newtheorem{definition}[theorem]{Definition}
\newtheorem{lemma}[theorem]{Lemma}

\newtheorem{proposition}{Proposition}
\newtheorem{example}{Example}

\title{A Non-Interventionist Approach 
to Causal Reasoning
based on Lewisian  Counterfactuals}

\begin{document}

\maketitle 
\begin{abstract}
We present a computationally grounded semantics for counterfactual
conditionals 
in which i) the
state in a model is decomposed into two elements:
a propositional valuation and a causal base
in propositional form
that
represents  
the causal
information available at the state; and ii) 
the 
comparative similarity relation between states is computed from the states' two components. 
We show that, by means of our semantics, 
we can elegantly  formalize 
the notion of actual
cause without recurring to the primitive notion of intervention. Furthermore,
we provide a   succinct
formulation of the model
checking problem for a language of counterfactual
conditionals in our semantics.
We show that this problem is PSPACE-complete
and provide a reduction
of it into QBF that can be used
for automatic verification
of causal properties.
\end{abstract}

\section{Introduction}%
\label{sec:intro}

The theory of counterfactual
conditionals is one of the cornerstones  of modern
analytic philosophy since the seminal
works
of Lewis 
\cite{LewisCounter} and 
Stalnaker 
\cite{StalnakerConditionals68}.
It has been recently applied
in the field of explainable
AI to explain the decisions and predictions 
of artificial intelligent  systems 
\cite{mittelstadt2019explaining,mothilal2020explaining,sokol2019counterfactual,DBLP:conf/aaai/KennyK21}. 
The theory of
counterfactual conditionals
is  intimaly related
to the theory of causation,
and the logic of counterfactual reasoning 
to the logic of  causal reasoning. 

As an alternative to Lewis' logic
of counterfactual conditionals, Halpern and Pearl 
\cite{GallesPearl1998,Halpern2000,HalpernPearl2005a}
have introduced 
a logic
of interventionist conditionals
as a special kind of counterfactual
conditionals
in which the antecedent of
the conditional is an intervention
on a causal model. 
Unlike Lewis who interprets 
his logic
of counterfactual
conditionals
by means of abstract comparative
similarity relations between possible
worlds, Halpern and Pearl 
interpret their logic by means of  a structural
equation model (SEM) semantics.
The fact that counterfactual
conditionals are more general
than interventionist conditionals 
is emphasized by Pearl 
 \cite{Pearl2009}
who introduced the notion of the three-layer `causal
ladder' (or hierarchy) in which counterfactuals
are at the top of 
of the hierarchy, interventions are in the middle
and mere associations are at the bottom layer. 
 Counterfactuals are placed at the top of the ladder
since they subsume interventions, in the sense that an interventional
question can be formulated as a special
kind of counterfactual question but not vice versa. 

The theory and the corresponding logic
of interventionist conditionals have become 
the dominant paradigm  in the field 
of formal causal reasoning 
in AI in the recent years, while
Lewisian conditionals are much less prominent. 
A variety of causal concepts
have been formalized using   interventionist conditionals 
including
actual cause
\cite{Halpern2016,DBLP:conf/aaai/Beckers21a,DBLP:conf/kr/Halpern08a,BeckersVENNEKENS},
NESS (Necessary Element of a Sufficient Set) cause \cite{DBLP:conf/aaai/Beckers21a,DBLP:conf/kr/Halpern08a}, 
contrastive cause \cite{DBLP:journals/ker/Miller21},
explanation \cite{HalpernPearl2005b,WoodwardBook2003,woodward2003explanatory},
responsibility and blame \cite{DBLP:journals/jair/ChocklerH04,DBLP:conf/aaai/HalpernK18,DBLP:conf/atal/AlechinaHL17}, 
discrimination \cite{DBLP:conf/aaai/ChocklerH22}
and 
harm \cite{DBLP:conf/nips/BeckersCH22}. 
Thus, the general impression we get from these  works is that the
primitive notion of intervention 
is necessary to define and formalize such causal concepts.
In this paper, we show that this impression is not well-founded. 
In particular, we prove that the notion of actual cause,
one of the central pillars in the modern  theory of causality, can be naturally and elegantly 
formalized 
in a language of counterfactual conditionals in Lewis' style
without recurring to the notion of intervention.

To obtain our result, we rely on
the computationally grounded 
semantics for causal reasoning recently proposed in \cite{DBLP:conf/ijcai/Lorini23,de2024model}. 
There is a crucial difference between
Lewis' original 
semantics
for  counterfactual conditionals
and the semantics on which we rely.
In the former, the notion of possible
state (or world) in a model 
is undecomposed 
and the 
comparative similarity 
relation between states
used to interpret counterfactuals 
is abstract. In the latter, a state is decomposed into two elements:
i) 
a propositional valuation, and ii)  a causal base
in propositional form
that 
represents the   causal
information available at the state. 
Moreover, the comparative similarity relation is grounded in  and computed  from the states' two components.
In this sense, it is a two-dimensional semantics for counterfactual conditionals. 
Specifically,
according to this 
semantics, 
a state $S' $
is considered at least as similar
to a state $S$
as a state $S''$
 if i) 
the causal
information the state $S'$
shares with  the state $S$
is 
at least as much as
the causal information the state 
$S'' $
shares with the state $S$,
and ii) $S''$ 
differs from $S$
with respect to the truth values of 
propositional atoms
at least as much as $S'$
differs from $S$. 

Our semantics offers greater
flexibility than the abstract Lewisian semantics
and allows us to give a precise interpretation of  Lewis' vague concept
of a
`small miracle' \cite{Lewis1979-LEWCDA}.
Lewis uses this concept to distinguish
backtracking from  
non-backtracking counterfactuals. 
 Roughly speaking, according to Lewis, in a
backtracking counterfactual only the propositional atoms representing 
the
initial
conditions can be 
changed to satisfy 
the antecedent of the 
conditional, while the causal laws are kept fixed.
On the contrary, in a non-backtracking counterfactual, the causal laws can  be changed
by imagining  
`small miracles'. 
According to 
the two-dimensional  semantics
we use, a `small miracle'
is nothing but a \emph{minimal change of
a causal base
that can possibly occur 
  to satisfy
the antecedent of a conditional}. 



The paper is structured as follows.
In Section \ref{sec:relwork} we discuss some 
work that is directly related
to our work. 
In Section  \ref{sec:framework}
we present the formal framework:
the two-dimensional semantics,
the language of counterfactual
conditionals
and its interpretation
over it,
and a list of interesting validities
for this language.
Section
\ref{sec:actcause}
presents the main conceptual
result
of the paper.
After  some formal
preliminaries 
introducing the notion
of equational state,
we prove a theorem
highlighting that  the notion
of actual cause, as  
defined
in \cite{DBLP:conf/ijcai/Halpern15}
using the notion
of intervention, 
can be equivalently
defined in our language
of Lewisian counterfactuals
  without recurring
to interventions. 
Section \ref{sec:mc}
is devoted to the computational aspects
of our novel semantic approach
to counterfactual conditionals. 
We provide a   succinct 
formulation of the model
checking problem for the language of counterfactual
conditionals in our semantics.
With `succinct' we mean that the model
with respect
to which a formula
has to be checked is not given explicitly
with its set of possible worlds and its comparative similarity relations, but it is  given in a compact form. 
We show that this problem is PSPACE-complete
and provide a reduction
of it into QBF that can be used
for automatic verification
of causal properties. 
As far as we know, 
nobody before us provided a succinct formulation
of the model checking problem for Lewis' logic
of counterfactual conditionals and
a tight complexity result for this problem. 

All proofs are given in the appendices.

\section{Related Work}\label{sec:relwork}

 The connection
between the  logic
of interventionist
conditionals
and Lewis' logic
of counterfactual conditionals
was studied in \cite{GallesPearl1998}
and more recently in \cite{DBLP:journals/mima/Zhang13}.
Galles \& Pearl show how a comparative similarity
relation between possible worlds 
can be computed by a means
of interventions: 
a first world is more similar
to a second world than a third world is if
it takes less local
interventions
to transform
the first
world
into the second
world
than to transform
the third world
into
the second world. 
As noticed by
Zhang,  the semantics
of counterfactual
conditionals
based  on selection functions
in Stalnaker's style
can also be reconstructed by means of interventions: the function
selects
for each intervention the solutions
of the underlying causal model
produced by it, 
as the closest worlds
to the actual one
relative to the intervention. 
Zhang studies the subclass
of causal
models,
the so-called
\emph{solution-conservative} causal
models, for which the 
principles 
of the logic
of interventionist conditionals
that 
correspond to the 
axioms
of Lewis' logic
of counterfactual
conditionals 
are valid.  
However, Galles \&  Pearl's and
Zhang's approach
is fundamentally different from
our approach. They focus on the logic
of interventionist conditionals and aim to elucidate the relation with Lewis' logic. We focus on counterfactual conditionals
and get rid of interventions. We show that the notion of actual
cause has a natural and elegant interpretation
in the logic of counterfactual conditionals
 that do not require the notion of intervention. 
 
A recent analysis of the distinction
between backtracking and 
non-backtracking counterfactuals
in an interventionist setting
was given in \cite{DBLP:conf/clear2/KugelgenMB23}. This semantic account of non-backtracking counterfactuals is fundamentally different from ours.
Following Pearl \cite{Pearl2009}, they make the concept of non-backtracking counterfactual
conditional coincide with the concept
of interventionist conditional
and the concept
of `small miracle' with the concept
of intervention. As pointed out above, our interpretation
of Lewis' concept
of a `small miracle'
does not rely
on the concept
of intervention but rather
on the concept
of minimal
change of a causal base. 

Alternative semantics for actual causality based on the situation calculus (SC) have also been proposed. Batusov and Soutchanski \cite{Batusov_Soutchanski_2018} formalize actual causality using atemporal SC action theories with sequential actions. Khan and Lespérance \cite{khanKnowingWhyDynamics2021} extend causal reasoning to epistemic contexts involving incomplete information and multiagent settings, analyzing how agents acquire knowledge of actual causes through actions and sensing.

 Last but not least, it is worth mentioning the work on the connection between counterfactuals and causal rules in the framework of causal calculus presented in \cite{DBLP:conf/kr/Bochman18,Bochman2021}. We share with Bochman and previous work in \cite{DBLP:conf/ijcai/Lorini23,de2024model}
the idea of expressing causal
information 
  through causal
rules
expressed in propositional form, 
as an alternative
to the SEM semantics 
of 
Halpern and Pearl 
and to the causal
team semantics
introduced in  \cite{DBLP:journals/jphil/BarberoS21}.


 \section{Formal Framework}
 \label{sec:framework}

In this section, we first present the two-dimensional
semantics
for counterfactual
conditionals.
Then, we introduce a language 
that supports
reasoning about
propositional facts,
information in a causal base
and counterfactuals.
We show how the language
can be interpreted using the two-dimensional semantics.
Finally,
we discuss some of its formal properties
in relation to Lewis' logic.

\subsection{Semantics  }\label{sec:causesemantics}
In 
\cite{DBLP:conf/ijcai/Lorini23}
a rule-based semantics for causal reasoning
is presented.
The main feature of the semantics
is its two-dimensional nature: one dimension representing the actual
environment, and the other dimension representing the causal information. 
In this section, we extend this semantics
with comparative similarity 
relations to be able to interpret
counterfactual conditionals. 

 Let $\Props$ be
 an infinite countable set of atomic propositions  whose elements are denoted $\prop, \propb, \ldots $
 We note $\langP (\Props)$,
 or simply $\langP $,
 the propositional language built from
 $\Props$.
 Elements of $\langP$
 are denoted $\omega, \omega', \ldots $
 Given $\omega \in \langP$,
 we note with $\Props(\omega)$
 the set of atomic propositions occurring
 in $\omega$. Moreover,
 if $\subbase \subseteq \langP$
 then $\Props(\subbase)=\bigcup_{\omega \in \subbase} \Props(\omega)$. 

The following definition introduces the concept
of state, namely, a causal base
supplemented with a propositional
valuation that is compatible with it. 
\begin{definition}[State]
\label{def:state}
A state is a pair $\state = \tuple{\base,\val}$,
where $\base \subseteq \langP$ is a causal  base, 
and $\val \subseteq \Props$ is a valuation
s.t. $\forall \omega  \in \base, \val \sat \omega $.
The set of all states is denoted by $\univCtx$.
A state $\state = \tuple{\base,\val}$
is said to be finite if both 
$\base$
and $\val$
are finite.
\end{definition}
The propositional  valuation
$\val$
represents the actual environment,
while
$\base $
represents the base of causal
information (viz.
the causal base). It
is assumed that the former
is compatible with the latter,
that is, if $\omega  $
is included in the  actual causal base
(i.e., $\omega  \in \base $)
then it should be true
in the actual environment 
(i.e., 
$\val \models \omega 
$). 
We let super- and subscripts to be inherited, e.g., $S^*$ always stands for $(C^*, V^*)$. 


A model is nothing but
a state supplemented with a set of states that includes it. 
 \begin{definition}[Model]\label{def:model}
A model is
a pair $ \tuple{\state,\ctx}$
such that $\state \in \ctx \subseteq \univCtx$.
The set of models is denoted $\Models$.
\end{definition} 
The component  $\ctx   $ 
is called   \emph{context}
  (or \emph{universe}) 
  of interpretation. 
  We call $ \tuple{\state,\univCtx}$
  a universal model (i.e., a model including all possible states). 
  For notational convenience, we simply write
  $\state $ instead 
  $ \tuple{\state,\univCtx}$
  to denote a universal model.

  Let us illustrate
the previous  notion
of model 
with the help of an example. 
\begin{example}[Videogame]\label{ex:videogame}
Consider a virtual character controlled by a video gamer using three keyboard keys. Each configuration of these keys corresponds to a specific causal base, which determines the action the virtual character will perform depending on which key is activated by the gamer. Assume that three actions are possible: ‘move forward’ ($\mathit{fo}$), ‘move backward’ ($\mathit{ba}$), and ‘jump’ ($\mathit{ju}$).
Suppose that:
\begin{itemize}
    \item[i)] the controls are configured such that activating key 1 ($\mathit{ac}_{1}$) causes the character to move forward; activating key 2 ($\mathit{ac}_{2}$) causes it to move backward; and activating key 3 ($\mathit{ac}_{3}$) causes it to jump;
      \item[ii)]  in the actual situation, no key is activated and the character remains stationary; 
  \item[iii)]  a hard constraint in the game prevents the 
gamer from
activating more than one key at the same time.
\end{itemize}

So, according to hypotheses i), ii) and iii),  we are in a  model 
 $ (\state_0 ,\ctx_0)$
 with $ \state_0 = \tuple{\base_0,\val_0}$
 such that
 \begin{align*}
& \base_0=\big\{
\mathit{ac}_{1} \rightarrow
\mathit{fo},
\mathit{ac}_{2}  \rightarrow \mathit{ba},
\mathit{ac}_{3} 
 \rightarrow 
 \mathit{ju}
\big\},\\
& \val_0= \emptyset ,\\
& \ctx_0=
\big\{ 
\tuple{\base',\val' }
\in \univCtx
\suchthat 
\val' \models 
\bigwedge_{
\substack{
x,y\in \{1,2,3\}
\suchthat \\ x\neq y
}
}
(\mathit{ac}_{x}
\limp \neg
\mathit{ac}_{y})
\big\}.
\end{align*}

\end{example}

We define the following comparative similarity relation
between states.
\begin{definition}[Similarity relation between states]\label{def:similarity_relation}
Let $\state = \tuple{\base,\val},\state' = \tuple{\base',\val' },
\state'' = \tuple{\base'',\val'' } \in \univCtx $.
We say that state $\state '$
is at least as similar 
to state $\state $ as state $\state ''$ is, denoted $\state  ''\preceq_{\state } \state '$,
if
\begin{align*}
 (\base  \cap \base '')\subseteq  (\base  \cap \base') \text{ and }
 (V \Delta V') \subseteq (V \Delta V''),
\end{align*}
where $\Delta$ stands for symmetric difference.

\end{definition}
According
to the previous definition,
state $\state '$
is at least as similar 
to state $\state $ as state $\state ''$ is
if i) the causal
information that $\state''$
shares with $\state$
is included in the causal
information that $\state'$
shares with $\state$,
and ii)
the environment of $\state''$
differs from the environment of $\state$
at least as much as the environment of $\state'$
differs from the environment 
of $\state$.
The reason why the similarity relation
uses  `set-inclusion' for the causal
  part 
and `symmetric difference' for the propositional  part
is that the \emph{causal} similarity  between two states
is determined by the information that is shared by their causal bases, while 
their \emph{propositional}  similarity  
is determined by the set of atomic propositions whose truth values are the same in their propositional valuations.



\subsection{Language}
The following definition introduces our modal
language for causal reasoning. 

\begin{definition}[Language]
	We structure the language in two layers:
	\begin{align*}
	& \langE & \isdef && \alpha & \coloncolonequals \prop 
	\mid  \top  
	\mid 
	\lnot\alpha
	\mid \alpha\land\alpha
	\mid \ltri\omega ,
	\\
	& \langDI & \isdef && \phi & \coloncolonequals \alpha
	\mid \lnot\phi
	\mid \phi\land\phi
	\mid \varphi \mid \condmodal{\varphi}{\varphi}, 
	\end{align*}
	where $\prop$ ranges over $\Props$
    and $\omega $
    over $\langP$.  
The boolean constructs $\bot$, $\lor$, $\limp$, and $\lequ$ are defined in the standard way as abbreviations. 
\end{definition}

We call
$\langE$ the language of causal information
and
$\langDI$ the language of
  causal counterfactual conditionals. 
Formula $\ltri\omega $
is  read 
``it is causally relevant that $\omega $''
or 
``$\omega $ is a causal information
of the actual state''.
Formula 
$ \condmodal{\varphi}{\psi}$
is read 
``if $\varphi$ were true, $\psi $ would be true''.
Its dual   $\phi \diamondRight \psi
:= \neg (\phi \boxRight \neg \psi)
$ is read ``if $\phi$ were true, $\psi$ might be false''.

The following definition introduces 
the satisfaction relation between 
models 
and
formulas of 
the language $\langDI$.  
(We omit semantic interpretations for the boolean connectives
$\neg, \wedge$ and for $\top$ since they are  defined in the usual way.)
\begin{definition}[Satisfaction relation]\label{semint}
\label{def:sat}%
Let 
$ \tuple{\state,\ctx} \in \Models $
with   $\state = \tuple{\base,\val}$. 
Then, (boolean cases are omitted)
\begin{align*}
  \tuple{\state,\ctx} & \sat \prop   
  && \text{ iff } &&\prop \in  \val ,
  \\
  \tuple{\state,\ctx} & \sat \ltri\omega
  && \text{ iff }  && \omega \in \base,
  \\
  \tuple{\state,\ctx} & \sat \condmodal{\varphi}{\psi}
  && \text{ iff } && \text{for all } \state' \in 
  \closest (\varphi {,}\state {, } \ctx  )
  ,  \\
  &&&&& \tuple{\state',\ctx} \sat \psi,
\end{align*}
where 
\begin{align*}
\closest (\varphi {,}\state {, } \ctx  )=&
\Big\{ \state ' \in \ctx : 
  \tuple{\state',\ctx} \models \varphi  \text{ and }\not \exists  \state '' \in \ctx \\
& \text{ such that  } 
  \tuple{\state'',\ctx}
  \models \varphi    \text{ and }\state' \prec_{\state}  \state'' \Big\},
\end{align*}
and 
$\state' \prec_{\state}  \state''
$ iff $\state' \preceq_{\state} \state''$
and $\state'' \not \preceq_{\state}  \state'$.
\end{definition}
The set
$\closest (\varphi {,}\state {, } \ctx  )$
is the set of $\varphi$-closest
states 
to state $\state$
relative to the context $\ctx$.

The formula 
$\ltri \omega$
has the expected set-theoretic
interpretation: 
it is causally relevant that $\omega $
iff the propositional formula $\omega$
is included in the actual causal base. 
The  counterfactual conditional
$\condmodal{\varphi}{\psi}$ holds 
at model $ \tuple{\state,\ctx}$
if $\varphi$ were true, 
$\psi$
would be true iff all $\varphi$-closest
states to state $\state$
relative to the context $\ctx$
satisfy $\psi$.

  Let us 
  go back
  to Example
  \ref{ex:videogame}
  of the videogame
  to illustrate
  the semantic
  interpretation
  of formulas. 
\begin{example}[Videogame continued]\label{ex:videogame2}
It
is easy to verify  that
at model
 $ (\state_0 ,\ctx_0)$
i)  if the gamer  activated 
 key $3$, the virtual character might jump, and ii)
 if the gamer activated  key
 $3$
 without changing the causal
 rule relating
key $3$
to the jumping action, the virtual
 character would jump, that is, 
  \begin{align*}
 (\state_0 ,\ctx_0)\models 
 (\mathit{ac}_{3}
 \diamondRight
 \mathit{ju})
 \wedge
\Big(\big(\condmodal{\mathit{ac}_{3}\wedge
 \ltri
(\mathit{ac}_{3} 
 \rightarrow 
 \mathit{ju})
 \big)}{ \mathit{ju}} \Big).
   \end{align*}

\end{example}

Recall that we write $\state  \models \phi$ instead of $ \tuple{\state,\univCtx}
\models \phi$ for notational convenience.
We say that a formula $\varphi\in \lang(\Props)$
is valid,
denoted by $\models \varphi $,
if $ \tuple{\state,\ctx} \models \varphi$
for every model $\tuple{\state,\ctx}\in \Models$. We say $\varphi$
is satisfiable
if $\neg \varphi$
is not valid.


\subsection{Some Properties}\label{sec:properties}


The following proposition highlights
some interesting properties 
of our counterfactual conditionals. 
\begin{proposition}\label{prop: validities}
    {
    Let $\phi,\psi \in \langDI$, $\omega \in \langP$ and $p \in \Props $}.
    We have the following validities:
\begin{align}
& \models \phi \boxRight \phi \label{validity: id}\\
& \models (\phi \boxRight \psi) \rightarrow (\phi \rightarrow \psi) \label{validity:WC} \\
    & \models \big( \phi \boxRight \chi \wedge \psi \boxRight \chi \big) \rightarrow (\phi \vee \psi) \boxRight \chi \label{validity: 3} \\
 &   \models \big( p   \wedge (\phi \boxRight \psi) \big) \rightarrow (\phi \wedge p  ) \boxRight \psi \label{validity:termRM1}\\ 
  &   \models \big( \neg p   \wedge (\phi \boxRight \psi) \big) \rightarrow (\phi \wedge \neg p   ) \boxRight \psi \label{validity:termRM2}\\ 
   &   \models \big( \ltri \omega   \wedge (\phi \boxRight \psi) \big) \rightarrow (\phi \wedge \ltri \omega    ) \boxRight \psi \label{validity:termRM3} \\
   & \models \ltri \omega \boxRight \omega \label{validity: interaction} 
\end{align}
\end{proposition}
The first three validities can be proven straightforwardly.
The validity (\ref{validity: id}) is standard in conditional logics.
The validity  (\ref{validity:WC}) is called \emph{weak centering} in the literature \cite{LewisCounter}. The name comes from its semantic condition, namely if $\tuple{\state,\ctx} \models \phi$, then $\state  \in \closest(\phi, \state,\ctx)$.
However, the property \emph{strong centering}, i.e., if $\tuple{\state,\ctx}  \models \phi$ then $\{\state \} = \closest(\phi, \state , \ctx)$,  does not hold. 
A counterexample would be
\begin{align*}
    \ctx = \{\state ,  \state'\}
    \text{ with  } \state = \{ \emptyset, \{p\} \} \text{ and } \state' = \{ \{p\}, \{p\} \}.
\end{align*}
We have $\{\state \} \neq \closest(p, \state, \ctx) = \ctx$, albeit $(\state , \ctx) \models p$.
The validity  (\ref{validity: 3}) comes from the fact that
the comparative
similarity
relation $\preceq_\state$ of Definition 
\ref{def:similarity_relation}
is a partial preorder.

The validities (\ref{validity:termRM1}), (\ref{validity:termRM2}) and  (\ref{validity:termRM3}) are of particular interest since they highlight
the interaction between counterfactual
conditionals, propositional atoms
and causal information. 
If $\preceq_\state$ were a total
preorder,
the formula  
\begin{align*}
\big( (\phi \boxRight \psi) \wedge (\phi \diamondRight \chi) \big) \rightarrow (\phi \wedge \chi) \boxRight \psi    
\end{align*}
 would be valid.
This  formula is an axiom of Lewis' V-logics
that relates to many axioms/postulates in other fields, e.g., the last postulate in AGM theory \cite{alchourron1985logic}, and \emph{rational monotonicity} (RM) in non-monotonic reasoning \cite{kraus1990nonmonotonic}.
Since $\preceq_\state$ is not total, RM is not valid here.
Nevertheless, the validities (\ref{validity:termRM1}), (\ref{validity:termRM2}) and   (\ref{validity:termRM3})  indicate that our semantics is monotonic under cumulation of true propositional  atoms and their negation,    and of actual  causal information.
Finally, the validity (\ref{validity: interaction}) highlights  the interaction between causal information and counterfactual conditionals, and comes from the validity (\ref{validity: id}) and the validity of $\ltri \omega \rightarrow \omega$.

\section{Actual Cause}\label{sec:actcause}

In this section,
we turn to actual cause. 
We first provide
some preliminary notions,
the notion
of equational state
and the notion
of intervention,  that are needed
to define
 actual cause in Halpern \&  Pearl's sense. We focus 
on the most recent interventionist 
definition
of actual
cause
given in \cite{DBLP:conf/kr/Halpern08a}. 
The section culminates  with a theorem
showing that 
 Halpern's 
notion of actual
cause
can be equivalently
formulated
in our language of counterfactual
conditionals
without  interventions. 

\subsection{Equational States}\label{sec:eqsemantics}

We  consider a subclass
of states in which, in line with
the structural equational modeling (SEM)
 approach to causality, 
causal information is represented in equational form.

An equational formula
for a proposition 
$\prop$
is a propositional formula
of the form 
$ p  \leftrightarrow \omega$
which  unambiguously specifies the truth value of $\prop$ using
a propositional formula
$\omega$
made of propositions  other than $\prop$,
with $ \leftrightarrow$
the usual biconditional 
boolean connective ``if and only if''. 
We note $  \langequat$ the corresponding set of equational formulas:
\begin{align*}
 \langequat=
 \Big\{ 
 p  \leftrightarrow \omega
 :
 p \in \Props, 
 \omega \in \langP ; 
 \text{ and } p \not \in \Props(\omega)
 \Big\}  .
 \end{align*}
For every $\prop \in \Props$,
$\langequat (\prop)$
is the set of equational formulas for $\prop$. 
For notational convenience,
elements of $ \langequat$
are also  denoted $\epsilon, \epsilon', \ldots$

An equational state is a special kind of state
whose  causal base 
is 
a finite  set of equational formulas.
\begin{definition}[Equational state] \label{def:eqState}
An equational
state is a state $\state = \tuple{\base,\val}$,
with  $\base \subseteq  \langequat $ finite,  and 
such that
\begin{align*}
  \forall \prop \in \Props, 
     \forall \prop\leftrightarrow\omega, \prop\leftrightarrow\omega' \in \base,
       \omega=\omega'.   
\end{align*}
The set of equational  states 
is denoted  by $\univCtxEq$.
\end{definition}
According to the previous definition, the  causal
base of an equational state should contain  at most one equational
     formula for each atomic proposition.

From an  equational state, it is straightforward
to extract a
  a set of endogenous variables and a set of 
exogenous ones.
A variable is endogenous if there is an equational formula 
for it in the actual causal base, 
it  is exogenous if it appears
in the actual   causal
base but there is no equational formula for it.

\begin{definition}[Exogenous and endogenous variables]
    Let 
    $\state= \tuple{\base,\val}$ be an equational state. 
Its set of exogenous variables
$   \exgfunct(\state ) $ and
its  set of  endogenous variables $   \engfunct(\state )$
are defined, as follows:
\begin{align*}
     \engfunct(\state  )  = &\big\{\prop \in 
\Props(\base)   : 
\exists \omega  \in 
 \langP \big(\Props  \setminus \{\prop \} \big)
 \text{ such that } \\
 &
  \prop   \leftrightarrow \omega \in 
   \base
   \big\} ,\\
     \exgfunct(\state ) = & \Props (\base ) \setminus  \engfunct(\state  ). 
\end{align*}
\end{definition}

From an equational state 
it
is also possible 
to extract its graphical counterpart. 
Specifically, given
an equational  state 
$\state = \tuple{\base,\val}\in \univCtxEq$, 
    we can extract the causal graph
$\causegraph_{\state }=\big(\nodeset_\state , \graphneigh_\state  \big)$
with $\nodeset_\state = \Props^+ (\base )
$
and 
where the \emph{causal parent} function
$ \graphneigh_\state \colon \nodeset_\state \longrightarrow 2^{\nodeset} $
is defined as follows, for every $ \prop \in \nodeset_\state $:
\begin{align*}
&(i) \ 
 \graphneigh_\state (\prop  )= \Props^+(\omega)
 \text{ if }
  \prop   \leftrightarrow \omega \in 
   \base ,\\
&(ii) \  \graphneigh_\state (\prop )= \emptyset \text{  if }
\langequat (\prop) \cap   \base =\emptyset,
\end{align*}
where $\Props^+(\omega)=\Props (\omega)\cup \{ \top \}$
if $\top $
occurs in $\omega$,
$\Props^+(\omega)=\Props(\omega)$
if $\top $
does not occur in $\omega$,
and $\Props^+(\base)= \bigcup_{\omega\in \base } \Props^+(\omega )$. Note that
if 
$\prop\in \nodeset_\state $
then,
$ \graphneigh_\state (\prop )= \emptyset $
iff $\prop \in  \exgfunct(\state ) $.

The following example is a classic in the literature on formal models of causality. We use it to illustrate the previous definition. 
\begin{example}\label{eg:rock-throwing}
 Suzy and Billy decide to throw a rock simultaneously, aiming at the bottle. Suzy is a bit faster, so her rock breaks the bottle, not Billy’s. However, Billy is just as accurate as Suzy: had she not thrown, Billy’s rock would have shattered the bottle shortly after. This leads to the following causal structure: i) Suzy throws her rock (st) iff she decides to do so (sd), ii) Billy throws his rock (bt) iff he decides to do so
(bd), iii) Suzy hits the bottle (sh) if and only if she throws her rock (st), iv) Billy hits the bottle (bh) if and only if he throws his rock (bt) while Suzy does not hit the bottle (¬sh), v) the bottle is shattered (bs) if and only if either Billy or Suzy hits it.
The actual state $\state_0 = \tuple{\base_0, \val_0}$ is described
as follows:
\begin{align*}
    \base_0=&\big\{ 
    \mathit{st} \leftrightarrow \mathit{sd},
        \mathit{bt} \leftrightarrow \mathit{bd},\\
    &    \mathit{sh} \lequ  \mathit{st},
    \mathit{bh} \lequ  (\mathit{bt} \wedge \neg  \mathit{sh}),
    \mathit{bs} \lequ (\mathit{sh} \vee \mathit{bh})
   \big\}, \\
   \val_0 = & \{sd, bd, st, bt, sh, bs\}.
\end{align*}
The causal graph extracted
from it is given in Figure
\ref{fig:example_suzy}. 
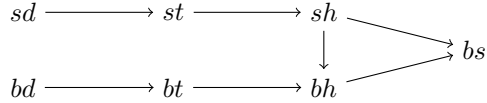
\begin{figure}[ht!]
	\centering
	\begin{tikzpicture}

        \node(SD) at (3,1){$sd$};
		\node(ST) at (5,1) {$st$};
		\node(SH) at (7,1) {$sh$};
		\node(BD) at (3,0){$bd$};
		\node(BT) at (5,0) {$bt$};
		\node(BH) at (7,0) {$bh$};
		\node(BS) at (9,0.5) {$bs$};

		\draw[->] (SD) -- (ST);
		\draw[->] (ST) -- (SH);
		\draw[->] (SH) -- (BS);
		\draw[->] (SH) -- (BH);

		\draw[->] (BD) -- (BT);
		\draw[->] (BT) -- (BH);
		\draw[->] (BH) -- (BS);

	\end{tikzpicture}
	\caption{Causal graph }
	\label{fig:example_suzy}
\end{figure}

\end{example}

 \subsection{Interventions}

 
We
conceive 
an  intervention
as  a possibly empty  finite set of 
equational 
formulas
of type
$\locint{\prop}{\top}$
or 
$\locint{\prop}{\bot}$
with at most one equational formula
for each variable. 
We define the set of interventions
as follows: 
\begin{align*}
\Interv{ }  =&  \big\{   \{ 
\locint{\prop_1}{\tau_1}  ,
\ldots, \locint{\prop_k}{\tau_k} \} :
\forall 1 \leq k',k'' \leq k,\\
&\text{ if } k'\neq k'' 
 \text{ then }
\prop_{k'}\neq 
\prop_{k'' } 
 \text{ and }\tau_1, \ldots,
\tau_k \in \{\top, \bot \}
\big\}. 
\end{align*}
Elements of $\Interv{ }$ are denoted $\globint, \globint', \dots $.
 Given $\globint \in \Interv{ } $,
 let
\begin{align*}
\conjfunct{\globint }
 =_{\mathit{def}} 
\bigwedge_{\locint{\prop}{\top  } \in \globint } \prop  \wedge 
\bigwedge_{\locint{\prop}{\bot  } \in \globint } \neg \prop.
\end{align*}

For every 
finite set of atomic propositions $\ArbProps \subseteq \Props $,
we note  $\Interv{ \ArbProps }$
the set of interventions for 
$ \ArbProps$, that is, 
\begin{align*}
    \Interv{ \ArbProps } =& \big\{ \globint \in 
    \Interv{ } 
    : (\forall \prop \in \ArbProps, \locint{\prop}{\top }\in
    \globint\text{ or }
    \locint{\prop}{\bot } \in \globint) \\
    & \text{ and } (\forall \prop \not  \in \ArbProps, \locint{\prop}{\top } \not \in
    \globint\text{ and  }
    \locint{\prop}{\bot }\not  \in \globint) 
\big\}.
\end{align*}


From a semantic point of view, an intervention  $\{ 
\locint{\prop_1}{\tau_1}  ,
\ldots, \locint{\prop_k}{\tau_k} \}$
replaces  any equational formula 
for $\prop_{k'}$
with $1 \leq k'\leq k$
in a causal base by the  equational formula
$\locint{\prop_{k'}}{\tau_{k'}}$.
Following this idea, 
the following definition introduces the notion 
of causal compatibility post intervention. 
\begin{definition}[Causal compatibility post intervention]\label{DefAlternative2}
Let $\globint \in  \Interv{ } $.
We define
$\relstateinterv{\globint }$ to be  the binary relation on the set 
of states $\univCtx   $
such that, for every $\state = \tuple{\base,\val},
\state' = \tuple{\base',\val' }\in \univCtx $: 
\begin{align*}
&\state  \relstateinterv{\globint } \state ' \text{ if and only if }\base'= 
\big( 
\base \setminus 
\bigcup_{ \prop \in \Props (\globint) }
\langequat (\prop) 
 \big)   \cup \globint . 
\end{align*}
\end{definition} 
$\state  \relstateinterv{\globint } \state '$
means that state $\state'= \tuple{\base',\val'}$ is compatible
with  state $\state = \tuple{\base,\val}$
after the occurrence of the  intervention  $\globint $.
Specifically, 
the latter is the case if
the causal base 
$  \base'$
is the 
result
of the following replacement
operation
applied to the causal base $  \base$: first of all 
remove from 
$  \base $
all  equational formulas
for the propositions 
on which we intervene through  $\globint $, 
and then add to the resulting causal  base 
 all  equational formulas
included in     $ \globint $.
Note that if $\state \in \univCtxEq$ and $\state  \relstateinterv{\globint } \state '$
then $\state' \in \univCtxEq$.
This means that 
intervening on an equational state
results in an equational state.

\subsection{Formalization of Actual Cause}\label{sec: actual cause formalization}

In this section, we recall  the  definition 
of actual cause 
given in  \cite{de2024model}. 
As shown by de Lima \&  Lorini,
under the assumption that the causal
graph induced by
the underlying  equational state 
is a DAG (directed acyclic graph)
their definition is equivalent to Halpern's definition given
in \cite{DBLP:conf/ijcai/Halpern15}.

Before defining actual
cause formally, some preliminary notation is needed. 
A  term is a 
conjunction of literals in which a propositional variable
can occur at most once. 
The set of terms is denoted by 
 $\Terms$ with elements  $\term, \term', \ldots$ 
 The set 
$\Terms_\ArbProps$
with $\ArbProps\subseteq \Props$
denotes the  set of terms
built from the variables in $\ArbProps $.
Given $\term , \term'\in \Terms$,
with a slight   abuse of notation,
we write $\term' \subseteq \term$
(resp. $\term' \subset \term$)
to mean that the set of literals
appearing in $\term'$
is a subset (resp. strict subset)  of the set of literals
appearing in $\term$.
Lastly, $\overline{\lambda}$ denotes the conjunction of 
the negations of  $\lambda$'s literals. 
That is, 
\begin{align*}
    \overline{\lambda} =_{\mathit{def}} \bigwedge_{p \subseteq \lambda} \neg p \wedge \bigwedge_{\neg p \subseteq \lambda} p.
\end{align*}




The definition below introduces the so-called 
{
``but''} condition. A term 
$\term$
    is a ``but'' condition
       for a propositional fact  $\omega$ at a state $\state$ if there exists an intervention on the endogenous variables in $\lambda$, along with another intervention that fixes the actual values of some endogenous variables not included in $\lambda$, such that, if the values of the exogenous variables remain unchanged, the formula $\varphi$ will necessarily be false after these interventions.
\begin{definition}[``But'' condition] \label{def:Butcause}
    Let $\state = \tuple{\base,\val} \in \univCtxEq $,
    $\term \in \Terms_{ \engfunct(\state ) }$
    and $\omega \in \langP$
    such that $\Props(\omega)\subseteq \Props(\base )$. 
    We say that $\term$
    is a ``but'' condition
    for $\omega$ at state $\state $,
    denoted by $\butcause(\state ,\lambda,\omega)$,
    if 
    \begin{align*}
  & \exists \globint \in 
      \Interv{ \Props(\lambda ) }, \exists 
    \ArbProps \subseteq 
\engfunct(\state ),
\exists \globint'  \in  \Interv{\ArbProps } 
 \text{ such that } \\
&         \ArbProps \cap \Props(\lambda )=\emptyset,
\state  \sat       \conjfunct{\globint' }
    \text{ and }\\
& \forall \state' \in \States ,
\text{ if } \state \relstateinterv{\globint\cup \globint'  }  \state'
\text{ and }\state'\sat \term_{\state}^{\mathit{exo}}
\text{ then  }
\state' 
    \sat \neg \omega     , 
\end{align*}
\begin{align*}
\term_{\state}^{\mathit{exo}}
=_{\mathit{def}}  \bigwedge_{\prop \in \exgfunct(\state ) \cap \val } \prop \wedge 
\bigwedge_{\prop \in \exgfunct(\state ) \setminus \val }\neg  \prop. 
\end{align*}
\end{definition}
 
The definition seems complicated, especially the $\globint '$ part
that consists in  fixing
the actual truth values
of some endogenous variables. 
The existential quantification
over such 
$\globint '$ is a core aspect
of Halpern's definition
of actual
cause.
This quantification is needed
to check the  absence of causal influence
from the other
variables on the produced effect. 
As we will show in Section \ref{sec: Reduc to Counterf},
this quantification is not needed when expressing actual cause through counterfactuals. 

We use 
the notion of ``but'' condition
to define the notion of actual cause below. 
Namely, $\term$
is an actual cause of $\omega $
if both $\term$
and $\omega$
are true, and   $\lambda $
is a \emph{minimal} ``but'' condition for $\omega $.

\begin{definition}[Actual cause]\label{def:ActualCause}
    Let $\state = \tuple{\base,\val} \in \univCtxEq $,
    $\term \in \Terms_{ \engfunct(\state ) }$
    and $\omega \in \langP$
    such that $\Props(\omega)\subseteq \Props(\base )$. 
    We say $\term $
    is an actual cause of  $\omega  $
    at state $\state $ 
    if:
\begin{align*}
    i) \ & \state \models  \term \wedge \omega , \\
    ii) \ &  \butcause(S,\lambda,\omega) \text{ holds}, \\
        iii) \ & 
        \forall \lambda' \subset \lambda,   \butcause(S, \lambda', \omega)
        \text{ does not hold} .
\end{align*}
\end{definition}
Let us emphasize again that,
as shown in \cite{de2024model}, 
the previous    definition
of actual cause is equivalent to 
Halpern's definition given in  \cite{DBLP:conf/ijcai/Halpern15}
when the causal
graph induced by the equational
state $\state$
is a DAG. 

Let us go back to Billy and Suzy's example. 

\begin{example}[Billy and Suzy continued] \label{ex: B&S actual cause}
We have that
$st $
is an actual
cause of $bs$ at state $\state_0$ in Example \ref{eg:rock-throwing}, while
$bt $ is not, for $\butcause(\state_0, bt, bs)$ does not hold.  
    


        
\end{example}

\subsection{Reduction to Counterfactuals} \label{sec: Reduc to Counterf}

In this section, 
we  are going to present 
the central
conceptual result of the paper:
a theorem
highlighting that actual cause can be expressed by means 
of counterfactual conditionals
\emph{without interventions}. 
The following Lemma
 \ref{lem: key}
is the key to prove it.

\begin{lemma}  \label{lem: key}
    Let $\state = \tuple{\base,\val } \in \univCtxEq $  such that 
 its causal graph
$\causegraph_{\state }$
is a DAG,
$\term \in \Terms_{ \engfunct(\state ) }$
and $\Props(\omega )  \subseteq  \Props(\base ) $. 
If 
$\state \models \lambda $,
then
$\butcause(\state,\lambda,\omega)$
if and only if
\begin{align*}
    S \models   
\bigvee_{  \lambda' \in 
\Terms_{\Props(\lambda )}  }
\big(
(\lambda' \wedge \lambda^{\exgfunct}_S) 
\diamondRight \neg \omega \big).
\end{align*}
\end{lemma}

The lemma states that the ``but'' condition can be captured
in terms of a might-conditional
under the assumption that 
the underlying causal graph is a DAG. In particular, 
under the assumption that
the causal graph induced by 
the state $\state$ is a DAG, 
$\lambda $
    is a ``but'' condition
    for
    the propositional fact $\omega$ at $\state $
    if and only if, 
    at $\state$
    there exists
    a term
    $\lambda'$
    sharing its propositions
    with $\lambda$
    such that if $\lambda'$
were true
and the exogenous variables had
their actual truth values,
$\omega$ might be false.

We sketch the proof idea
of the lemma here.
In the first glimpse, we must construct an intervention $E \cup E'$ witnessing $\butcause(S, \lambda, \omega)$ from some $S' \in \closest( \lambda' \wedge \lambda^{\exgfunct}_{\state}, S, \univCtx)$
with $S' \models \neg \omega$ and vice versa.
Apparently, 
a concern would be that the intervention does  not necessarily give rise to the  closest states to $S$, as required by $S'$.
Nonetheless, such a concern is unfounded: 
since $\omega $
is a propositional formula, its  truth value is only determined by the propositional valuation.
The 
causal base plays the role, together with the fact that the causal graph $G_S$ is a DAG, of ensuring that 
we can associate some $S'$ with 
the states resulting from the intervention $E \cup E'$, in such a way that they share the same 
propositional valuation.


We are now in a position
to show
the main result
of this section, namely the following 
Theorem 
\ref{thm:actual_cause_minimal_might}.  

\begin{theorem}
\label{thm:actual_cause_minimal_might}
    Let $\state = \tuple{\base, \val} \in \univCtxEq $ s.t.\ $\causegraph_{\state}$ is a DAG, $\term \in \Terms_{\engfunct(\state)}$ and $\Props(\omega) \subseteq \Props(\base)$. 
    Then, $\lambda$ is an actual cause of $\omega$ at $S$, if and only if 
    \begin{align*} 
    \state \models &  \lambda \wedge
        \big((\overline{\lambda} \wedge \lambda^{\exgfunct}_{\state}) \diamondRight \neg \omega\big)
    \wedge   \bigwedge_{
    \substack{
    \ArbProps \subset \Props(\lambda), \\
    \lambda' \in \Terms_{\ArbProps } } } \big((\lambda' \wedge \lambda^{\exgfunct}_{\state})  \boxRight \omega \big). 
         \end{align*}


\end{theorem}

 According to Theorem~\ref{thm:actual_cause_minimal_might}, under the assumption that the underlying causal graph is a DAG,
the notion of actual cause
can be captured by a combination of conditionals
and one might-conditional. 
In particular, 
under the assumption that
the causal graph induced by 
the state $\state$ is a DAG, 
$\lambda $
    is an actual
    cause of $\omega$ at $\state $
    if and only if
    at $\state$
    i) $\lambda $
    is true,
    ii)  if 
    the truth values
    of all variables
    in $\lambda$
    were changed
and the exogenous variables had
their actual truth values,
$\omega$ might be false,
and iii) for  every term
$\lambda'$
built from a strict
subset of
the set of propositions
in  $\lambda$, 
 if $\lambda' $
    were true 
and the exogenous variables had
their actual truth values,
$\omega$ would be true. 
Theorem~\ref{thm:actual_cause_minimal_might} highlights the main message
of our paper: \emph{actual cause
is definable using counterfactual conditionals  without having to resort to interventions}.

\begin{example}[Billy and Suzy revisited]
    We have that $st$ is  an actual cause of $bs$, for $\state_0 \models st \wedge (\neg st \wedge \lambda^{\exgfunct}_{\state}) \diamondRight \neg bs) \wedge \lambda^{\exgfunct}_{\state} \boxRight bs$, but  $bt$ is not, for $\state_0 \models (\neg bt \wedge \lambda^{\exgfunct}_{\state}) \boxRight bs $.
\end{example}

\section{Model Checking}\label{sec:mc}




In this section, we study the model checking problem in the defined framework.
To date, satisfiability checking received more attention in the literature on counterfactuals: a seminal paper~\cite{friedman1994complexity} established PSPACE-completeness for it in general (with a few exceptions for some properties of similarity ordering) and subsequent works proposed various decision procedures~\cite{GGOS2009TableauxPCL,LellmanPattinson2012SequentLewis,GirlandoNegriOlivetti2021LabelledPCL}.
At the same time,
using standard methods
in model checking \cite{GradelOtto}, 
it is straightforward 
to verify that model checking can be performed in PTIME if the whole model
is given \emph{explicitly} as input, 
including the set of possible states
and the comparative similarity
relations. However, explicit models  may be extremely large
and so unpractical. 


In our semantics,
model checking can be formulated in a succinct way
since the model does not need
to be given explicitly:
the set of possible states
and the comparative similarity relations  can be  computed \emph{ex post}. 
Specifically, following \cite{de2024model},  we define a succinct ``relativized'' version of model checking 
in which three elements are given as input: i) a formula
$\varphi$
of the language $\langDI$
to be checked,
ii) a finite vocabulary $\Gamma $
of propositional facts from which the context $ \univCtxRel{\Gamma} \eqdef  {\big\{ 
\state= \tuple{\base, \val} \in \univCtx \suchthat  \base \subseteq \Gamma \big\}} $ is defined, and iii)  a finite state $\state $ from  $\univCtxRel{\Gamma} $
with respect to which
the formula $\varphi$
is evaluated. 
The context $\univCtxRel{\Gamma} $
includes all states whose causal
bases are constructed from $\Gamma$.


\begin{framed}
\noindent\textbf{Model checking problem.} \\
Input: $\inputform \in \langDI$, finite $\Gamma \subset 2^{\langP}$, finite $\state \in \univCtxRel{\Gamma}$. \\
Output: \texttt{true} if $\tuple{\state, \univCtx^{\Gamma}} \models \inputform$, \texttt{false} otherwise.
\end{framed}

In the rest of this section,
we are going to show that this problem is PSPACE-complete by its polynomial reduction to the quantified Boolean Formula problem (QBF, see the definition in the supplementary material) and vice versa.

Let $\Gamma$ and $\inputform$ be given. The set of relevant atoms is defined as follows: $\Sigma = \Props(\Gamma) \cup \Props(\inputform)$. We can represent each state of $\univCtxRel{\Gamma}$ by $|\Gamma| + |\Sigma|$ bits, defining which facts from $\Gamma$ are present in the causal base and which relevant atoms are present in the valuation. Accordingly, we use sets of variables $\stateVarsi = {\{ \statevarbasei \mid \omega \in \Gamma \}} \cup \{ \statevarvali \mid \prop \in \Sigma \}$ for $i \in \Nats$ to represent the states in the QBF encoding.
Then, any state from $\univCtxRel{\Gamma}$ corresponds to some valuation on variables from $\stateVarsi$.

We define an encoding function $\qbfSat(\varphi,\stateVarsi)$, which  maps a subformula $\varphi$ of $\inputform$ (or of some formula in $\Gamma$) into an open QBF formula
satisfiable exactly by valuations on $\stateVarsi$ that correspond to states satisfying $\varphi$ (boolean cases are omitted):
\[\begin{array}{l}
\qbfSat(\prop, \stateVarsi) =
\statevarvali  \\[1pt]
\qbfSat(\ltri \omega, \stateVarsi) = \statevarbasei \\[1pt]
\qbfSat(\condmodal{\varphi_1}{\varphi_2}, \stateVarsi) = \forall \stateVars{\qbfindi+1}. \; \qbfState(\stateVars{\qbfindi+1}) \rightarrow \\
\hspace{45pt}(\qbfClosest(\varphi_1, \stateVarsi, \stateVars{\qbfindi+1}) \rightarrow \qbfSat(\varphi_2, \stateVars{\qbfindi+1})) \\
\end{array}\] 


Notice that for encoding quantification over states we need to use a set of variables $\stateVars{\qbfindi+1}$ different from $\stateVars{\qbfindi}$, and we need to check that causal base and the valuation given by choice of values of $\stateVars{\qbfindi+1}$ will be compatible (as required by Definition~\ref{def:state}). For this, we use  the
following predicate $\qbfState$: 
\[ \qbfState(\stateVarsi)  =  \bigwedge_{\omega \in \Gamma} (\statevarbasei \rightarrow \qbfSat(\omega, \stateVarsi)) . \]



Predicate $\qbfClosest$ encodes the definition of the closest state (Definition~\ref{def:sat}). However, this definition uses predicate $\qbfSat$ on $\varphi$ twice: to assert that given state satisfies $\varphi$ and that no closer state satisfies $\varphi$. To keep the encoding polynomial, we need to merge these two instances into one via standard Tseitins Tranformation~\cite{Tseitin1983} by introducing an extra quantifier:
\[
\begin{array}{l}
\qbfClosest(\varphi, \stateVarsi, \stateVarsii) = \forall \stateVarsiii.\; \forall r.\; \qbfState(\stateVarsiii) \rightarrow \\
\hspace{5mm} (\qbfSat(\varphi, \stateVarsiii) \leftrightarrow r) \rightarrow  ((\qbfEq(\stateVarsii, \stateVarsiii) \rightarrow r) \wedge \\
\hspace{37mm} (\qbfCloser(\stateVarsi, \stateVarsii, \stateVarsiii) \rightarrow
\neg r)).  
\end{array} \]

Here $k = \max({i, j}) + 1$ (to ensure that variables are different). Predicates $\qbfEq$ and $\qbfCloser$ encode equality and similarity of states from $\univCtxRel{\Gamma}$ directly by definitions. Full encodings are included in the supplementary material.

Notice that $|\qbfState(\stateVarsi)| = \O{\sum_{\omega \in \Gamma} |\omega|}$ since $|\qbfSat(\omega, \stateVarsi)| = \O{|\omega|}$ for $\omega \in \langP$, while $\qbfEq(\stateVarsi, \stateVarsii)$ and 
$\qbfCloser(\stateVarsi, \stateVarsii, \stateVarsiii)$ require to do $\O{|\Gamma| + |\Sigma|}$ checks on corresponding variables. So the predicate $\qbfSat$ makes a recursive call for each immediate subformula exactly once with an overhead at each step that is polynomial w.r.t. size of the input. Thus, we have a polynomial-size reduction to QBF, which immediately implies PSPACE-membership of the model checking problem.

For PSPACE-hardness we provide a reverse reduction (from QBF). It is based on the observation that for ${\prop, \prop' \in \Props \setminus V}$ there are exactly two states in ${\closest(\prop \vee \prop', (\emptyset,\val), \univCtxRel{\emptyset})}$, one satisfying $p$ and one not satisfying it (but satisfying $p'$), so we can use a counterfactual with ${(p \vee p')}$ in the antecedent to emulate boolean quantification over $p$.

\begin{theorem}
\label{th:model_checking_lower_bound}
The model checking problem is PSPACE-complete.
\end{theorem}

Our reduction of actual cause to counterfactuals in Theorem~\ref{thm:actual_cause_minimal_might} requires model checking with respect to the context $\univCtx$ that contains all states. In general, model checking w.r.t. $\univCtx$ can not be easily reduced to model checking w.r.t. a context $\univCtxRel{\Gamma}$ defined from  a finite vocabulary $\Gamma$ (we believe the former problem belongs to a higher complexity class). However, we can perform such a reduction in the special case when the input formula does not contain nested counterfactuals (which is the case  for Theorem~\ref{thm:actual_cause_minimal_might}).

\begin{lemma}
\label{lem:mc_relativized_reduction_unnested}
If $\inputform \in \langDI$ does not contain nested counterfactuals then
$\tuple{\tuple{\base, \val}, \univCtx} \models \inputform$ iff $\tuple{\tuple{\base, \val}, \univCtxRel{\Gamma}} \models \inputform$ for $\Gamma = {\base \cup \{ \omega : \ltri \omega \textit{ is a subformula of } \inputform\}}$.
\end{lemma}


Due to this reduction, we can employ our QBF encodings to check actual cause via Theorem~\ref{thm:actual_cause_minimal_might}. Moreover, although the last conjunction over $\lambda' \in \Terms_{\ArbProps}$ includes exponentially many conjuncts, we can 
obtain a polynomial encoding if we replace this conjunction with quantification over terms (which we can also naturally represent with boolean variables). 
With this modification (see details in the supplementary material) we can achieve polynomial QBF encoding with the depth of quantifier alteration equals $2$: 
$\exists \forall$ in the second conjunct and $\forall \exists$ in the third conjunct of the formula in Theorem~\ref{thm:actual_cause_minimal_might}.
In this sense, our encoding is ``close'' to being optimal, since the checking of actual cause was shown to be $\Sigma_2^P$-complete in \cite{eiter2002complexity}
and so  only requiring
$\exists \forall$
alternation. 

\section{Conclusion}

Let's take stock.
We have shown that the notion of intervention is not essential for the formalization of actual cause,
one of the central concepts in the theory of causality. This concept
can be captured by Lewisian counterfactual conditionals
once a two-dimensional semantics
distinguishing the propositional
level
from the causal level
is adopted. 
We have also shown that  
model checking 
for  the language
of counterfactual conditionals 
defined in this semantics
is PSPACE-complete
by means of its reduction
into QBF and vice versa.   

Our contribution has an impact at both the conceptual and computational level. On the conceptual side, we offer a general framework for unifying counterfactuals and actual cause. On the practical side, we provide a semantics for counterfactuals in which model checking can be formulated succinctly. This is useful in practice for the automatic verification of causal properties. 

Directions for future work are manifold. First, we plan to explore the proof-theoretic aspects of our logic of counterfactual conditionals. In Section \ref{sec:properties}, we only presented some interesting validities. We plan to develop a sound and complete axiomatization. Second, we plan to implement the QBF translation given in Section \ref{sec:mc}, in order to experimentally investigate the automated verification of causal properties—actual cause in particular—in terms of computation time. Third, we plan to extend our analysis based on Lewisian counterfactuals to other notions of cause, with special attention to Wright’s notion of NESS cause \cite{Wright1988}. Finally, we plan to investigate the relationship between our counterfactual atemporal approach to actual cause and recent work on temporal causal reasoning \cite{Gladyshev2025}. To this aim, we will extend our framework with an $\mathsf{LTL}$ temporal component, in order to account for temporal information in a causal base as well as counterfactual reasoning about temporal facts.



\section*{Acknowledgments}

This work  is supported by the ANR projects EpiRL (grant number ANR-22-CE23-0029) and ALoRS (grant number ANR-21-CE23-0018-01). 

\bibliography{ArXiv}

\begin{thebibliography}{10}

\bibitem{alchourron1985logic}
Carlos~E. Alchourr{\'o}n, Peter G{\"a}rdenfors, and David Makinson.
\newblock On the logic of theory change: Partial meet contraction and revision functions.
\newblock {\em The journal of symbolic logic}, 50(2):510--530, 1985.

\bibitem{DBLP:conf/atal/AlechinaHL17}
Natasha Alechina, Joseph~Y. Halpern, and Brian Logan.
\newblock Causality, responsibility and blame in team plans.
\newblock In {\em Proceedings of the 16th Conference on Autonomous Agents and MultiAgent Systems (AAMAS 2017)}, pages 1091--1099. {ACM}, 2017.

\bibitem{DBLP:journals/jphil/BarberoS21}
Fausto Barbero and Gabriel Sandu.
\newblock Team semantics for interventionist counterfactuals: Observations vs. interventions.
\newblock {\em Journal of Philosophical Logic}, 50(3):471--521, 2021.

\bibitem{Batusov_Soutchanski_2018}
Vitaliy Batusov and Mikhail Soutchanski.
\newblock Situation calculus semantics for actual causality.
\newblock pages 1744--1752, 2018.

\bibitem{DBLP:conf/aaai/Beckers21a}
Sander Beckers.
\newblock The counterfactual {NESS} definition of causation.
\newblock In {\em Proceedings of the Thirty-Fifth {AAAI} Conference on Artificial Intelligence (AAAI-21)}, pages 6210--6217. {AAAI} Press, 2021.

\bibitem{DBLP:conf/nips/BeckersCH22}
Sander Beckers, Hana Chockler, and Joseph~Y. Halpern.
\newblock A causal analysis of harm.
\newblock In Sanmi Koyejo, S.~Mohamed, A.~Agarwal, Danielle Belgrave, K.~Cho, and A.~Oh, editors, {\em Advances in Neural Information Processing Systems 35: Annual Conference on Neural Information Processing Systems 2022, NeurIPS 2022, New Orleans, LA, USA, November 28 - December 9, 2022}, 2022.

\bibitem{BeckersVENNEKENS}
Sander Beckers and Joost Vennekens.
\newblock The transitivity and asymmetry of actual causation.
\newblock {\em Ergo}, 4(1), 2017.

\bibitem{DBLP:conf/kr/Bochman18}
Alexander Bochman.
\newblock On laws and counterfactuals in causal reasoning.
\newblock In {\em Proceedings of the Sixteenth International Conference on Principles of Knowledge Representation and Reasoning (KR 2018)}, pages 494--503. {AAAI} Press, 2018.

\bibitem{Bochman2021}
Alexander Bochman.
\newblock {\em A Logical Theory of Causality}.
\newblock MIT Press, 2021.

\bibitem{DBLP:journals/jair/ChocklerH04}
Hana Chockler and Joseph~Y. Halpern.
\newblock Responsibility and blame: {A} structural-model approach.
\newblock {\em Journal of Artificial Intelligence Research}, 22:93--115, 2004.

\bibitem{DBLP:conf/aaai/ChocklerH22}
Hana Chockler and Joseph~Y. Halpern.
\newblock On testing for discrimination using causal models.
\newblock In {\em Proceedings of the Thirty-Sixth {AAAI} Conference on Artificial Intelligence (AAAI-22)}, pages 5548--5555. {AAAI} Press, 2022.

\bibitem{de2024model}
Tiago de~Lima and Emiliano Lorini.
\newblock Model checking causality.
\newblock In {\em Proceedings of the 33rd International Joint Conference on Artificial Intelligence (IJCAI 2024)}, pages 3324--3332. ijcai.org, 2024.

\bibitem{eiter2002complexity}
Thomas Eiter and Thomas Lukasiewicz.
\newblock Complexity results for structure-based causality.
\newblock {\em Artif. Intell.}, 142(1):53--89, 2002.

\bibitem{friedman1994complexity}
Nir Friedman and Joseph~Y. Halpern.
\newblock On the complexity of conditional logics.
\newblock In {\em Principles of Knowledge Representation and Reasoning}, pages 202--213. Elsevier, 1994.

\bibitem{GallesPearl1998}
David Galles and Judea Pearl.
\newblock An axiomatic characterization of causal counterfactuals.
\newblock {\em Foundation of Science}, 3(1):151--182, 1998.

\bibitem{GGOS2009TableauxPCL}
Laura Giordano, Valentina Gliozzi, Nicola Olivetti, and Camilla Schwind.
\newblock Tableau calculus for preference-based conditional logics: Pcl and its extensions.
\newblock {\em ACM Trans. Comput. Logic}, 10(3), April 2009.

\bibitem{GirlandoNegriOlivetti2021LabelledPCL}
Marianna Girlando, Sara Negri, and Nicola Olivetti.
\newblock Uniform labelled calculi for preferential conditional logics based on neighbourhood semantics.
\newblock {\em Journal of Logic and Computation}, 31(3):947--997, 04 2021.

\bibitem{GradelOtto}
Erich Gr\"{a}del and Martin Otto.
\newblock On logics with two variables.
\newblock {\em Theoretical Computer Science}, 224:73--113, 1999.

\bibitem{Halpern2000}
Joseph~Y. Halpern.
\newblock Axiomatizing causal reasoning.
\newblock {\em Journal of Artificial Intelligence Research}, 12:317--337, 2000.

\bibitem{DBLP:conf/kr/Halpern08a}
Joseph~Y. Halpern.
\newblock Defaults and normality in causal structures.
\newblock In G.~Brewka and J.~Lang, editors, {\em Principles of Knowledge Representation and Reasoning: Proceedings of the Eleventh International Conference (KR 2008)}, pages 198--208. {AAAI} Press, 2008.

\bibitem{DBLP:conf/ijcai/Halpern15}
Joseph~Y. Halpern.
\newblock A modification of the {H}alpern-{P}earl definition of causality.
\newblock In {\em Proceedings of the Twenty-Fourth International Joint Conference on Artificial Intelligence (IJCAI 2015)}, pages 3022--3033. {AAAI} Press, 2015.

\bibitem{Halpern2016}
Joseph~Y. Halpern.
\newblock {\em Actual causality}.
\newblock MIT Press, 2016.

\bibitem{DBLP:conf/aaai/HalpernK18}
Joseph~Y. Halpern and M.~Kleiman{-}Weiner.
\newblock Towards formal definitions of blameworthiness, intention, and moral responsibility.
\newblock In {\em Proceedings of the Thirty-Second {AAAI} Conference on Artificial Intelligence, (AAAI-18)}, pages 1853--1860. {AAAI} Press, 2018.

\bibitem{HalpernPearl2005a}
Joseph~Y. Halpern and Judea Pearl.
\newblock Causes and explanations: a structural-model approach. {P}art {I}: Causes.
\newblock {\em British Journal for Philosophy of Science}, 56(4):843--887, 2005.

\bibitem{HalpernPearl2005b}
Joseph~Y. Halpern and Judea Pearl.
\newblock Causes and explanations: a structural-model approach. {P}art {II}: Explanations.
\newblock {\em British Journal for Philosophy of Science}, 56(4):889--911, 2005.

\bibitem{DBLP:conf/aaai/KennyK21}
Eoin~M. Kenny and Mark~T. Keane.
\newblock On generating plausible counterfactual and semi-factual explanations for deep learning.
\newblock In {\em Proceedings of the Thirty-Fifth {AAAI} Conference on Artificial Intelligence (AAAI 2021)}, pages 11575--11585. {AAAI} Press, 2021.

\bibitem{khanKnowingWhyDynamics2021}
Shakil~M. Khan and Yves Lesp{\'e}rance.
\newblock Knowing why --- on the dynamics of knowledge about actual causes in the situation calculus.
\newblock In {\em Proceedings of the 20th International Conference on Autonomous Agents and Multiagent Systems (AAMAS 2021)}, pages 701--709. International Foundation for Autonomous Agents and Multiagent Systems (IFAAMAS), 2021.

\bibitem{kraus1990nonmonotonic}
Sarit Kraus, Daniel Lehmann, and Menachem Magidor.
\newblock Nonmonotonic reasoning, preferential models and cumulative logics.
\newblock {\em Artificial intelligence}, 44(1-2):167--207, 1990.

\bibitem{LellmanPattinson2012SequentLewis}
Bj{\"o}rn Lellmann and Dirk Pattinson.
\newblock Sequent systems for lewis' conditional logics.
\newblock In Luis~Fari{\~{n}}as del Cerro, Andreas Herzig, and J{\'e}r{\^o}me Mengin, editors, {\em Logics in Artificial Intelligence}, pages 320--332, 2012.

\bibitem{LewisCounter}
David~K. Lewis.
\newblock {\em Counterfactuals}.
\newblock Harvard University Press, 1973.

\bibitem{Lewis1979-LEWCDA}
David~K. Lewis.
\newblock Counterfactual dependence and time's arrow.
\newblock {\em No\^{u}s}, 13(4):455--476, 1979.

\bibitem{DBLP:conf/ijcai/Lorini23}
Emiliano Lorini.
\newblock A rule-based modal view of causal reasoning.
\newblock In {\em Proceedings of the 32nd International Joint Conference on Artificial Intelligence (IJCAI 2023)}, pages 3286--3295. ijcai.org, 2023.

\bibitem{DBLP:journals/ker/Miller21}
Tim Miller.
\newblock Contrastive explanation: a structural-model approach.
\newblock {\em The Knowledge Engineering Review}, 36:e14, 2021.

\bibitem{mittelstadt2019explaining}
Brent Mittelstadt, Chris Russell, and Sandra Wachter.
\newblock Explaining explanations in \text{AI}.
\newblock In {\em Proceedings of the 2019 conference on Fairness, Accountability, and Transparency}, pages 279--288, 2019.

\bibitem{mothilal2020explaining}
Ramaravind~K. Mothilal, Amit Sharma, and Chenhao Tan.
\newblock Explaining machine learning classifiers through diverse counterfactual explanations.
\newblock In {\em Proceedings of the 2020 Conference on Fairness, Accountability, and Transparency}, FAT* '20, page 607–617, New York, NY, USA, 2020. Association for Computing Machinery.

\bibitem{Pearl2009}
Judea Pearl.
\newblock {\em Causality: Models, Reasoning and Inference}.
\newblock Cambridge University Press, 2009.

\bibitem{sokol2019counterfactual}
Kacper Sokol and Peter Flach.
\newblock Counterfactual explanations of machine learning predictions: Opportunities and challenges for ai safety.
\newblock In {\em Proceedings of the AAAI Workshop on Artificial Intelligence Safety 2019}, volume 2301 of {\em CEUR Workshop Proceedings}. CEUR Workshop Proceedings, January 2019.
\newblock 2019 AAAI Workshop on Artificial Intelligence Safety, SafeAI 2019 ; Conference date: 27-01-2019.

\bibitem{StalnakerConditionals68}
Robert Stalnaker.
\newblock A theory of conditionals.
\newblock In N.~Rescher, editor, {\em Studies in Logical Theory}, pages 28--45. Oxford University Press, 1968.

\bibitem{DBLP:conf/clear2/KugelgenMB23}
Julius von K{\"u}gelgen, Abdirisak Mohamed, and Sander Beckers.
\newblock Backtracking counterfactuals.
\newblock In {\em Conference on Causal Learning and Reasoning, CLeaR 2023, 11-14 April 2023, Amazon Development Center, T{\"{u}}bingen, Germany, April 11-14, 2023}, volume 213 of {\em Proceedings of Machine Learning Research}, pages 177--196. {PMLR}, 2023.

\bibitem{WoodwardBook2003}
James Woodward.
\newblock {\em Making Things Happen: a Theory of Causal Explanation}.
\newblock Oxford University Press, 2003.

\bibitem{woodward2003explanatory}
James Woodward and Christopher Hitchcock.
\newblock Explanatory generalizations, part i: A counterfactual account.
\newblock {\em No{\^u}s}, 37(1):1--24, 2003.

\bibitem{wrathallCompleteSetsPolynomialtime1976}
Celia Wrathall.
\newblock Complete sets and the polynomial-time hierarchy.
\newblock {\em Theoretical Computer Science}, 3(1):23--33, October 1976.

\bibitem{Wright1988}
Richard~W. Wright.
\newblock Causation, responsibility, risk, probability, naked statistics, and proof: Pruning the bramble bush by clarifying the concepts.
\newblock {\em Iowa Law Review}, 73:1001--1077, 1988.

\bibitem{DBLP:journals/mima/Zhang13}
Jiji Zhang.
\newblock A {L}ewisian logic of causal counterfactuals.
\newblock {\em Minds and Machines}, 23(1):77--93, 2013.

\end{thebibliography}
\bibliographystyle{plain}

\newpage
\appendix

\begin{center}
    {\huge \textbf{Appendices}}
\end{center}

\section{Proof of Proposition \ref{prop: validities}}
\begin{proof}
We only show the validities (\ref{validity:termRM1}), (\ref{validity:termRM2}) and (\ref{validity:termRM3}), since as mentioned the others are straightforward to prove.

Let $\gamma$ denote any formula of the form $p$, $\neg p$ or $\ltri \omega$.
    Suppose towards a contradiction a model $\tuple{\state, \ctx}$, s.t. $\state \in \ctx \subseteq \univCtx$, where $\state \models \gamma \wedge (\phi \boxRight \psi)$ but $\state \not\models (\phi \wedge \gamma) \boxRight \psi$. The latter means $\exists S' \in \closest( \phi \wedge \gamma, \state, \ctx) $ s.t. $(S', \ctx) \models \neg \psi$. 
    Now clearly $S' \notin \closest( \phi, \state,U)$, for otherwise $\state \not\models \phi \boxRight \psi$ a contradiction. Therefore $\exists S'' \prec_S S'$ with $S'' \models \phi$. But since $S \models \gamma$ and $S' \models \gamma$, it has to be $S'' \models \gamma$ as well, otherwise either $V \Delta V'' \not\subset V \Delta V'$ or $C \cap C'' \not\supset C \cap C'$. However, this means $S' \notin \closest(\phi \wedge \gamma, \state, \ctx)$ in the first place, a desired contradiction.
\end{proof}

\section{Reduction to Counterfactuals}
\subsection{Some Preliminaries}

In this section we recall some results that were proved in \cite{de2024model} that will be needed for the proof of Theorem \ref{thm:actual_cause_minimal_might}.

\begin{proposition}[\cite{de2024model}]\label{prop:uniquesol}
    Let $\state = \tuple{\base,\val } \in \univCtxEq $  such that 
 its causal graph
$\causegraph_{\state }$
is a DAG
and let $   \globint \in 
      \Interv{ \ArbProps  }$
      for some $\ArbProps \subseteq   \engfunct(\state )$. 
Then,
there is a unique $\state' = \tuple{\base',\val' }$
such that 
     \begin{align*}
\state
\relstateinterv{\globint  }  
\state'
\end{align*}
and 
     \begin{align*}
& \val \cap \Big( \exgfunct(\state ) 
\cup\big( \Props \setminus \Props (\base )\big) 
\Big) 
=\val' \cap \Big( \exgfunct (\state ) 
\cup\big( \Props \setminus \Props (\base )\big) 
\Big) . 
\end{align*}
We denote with 
$\state^{\globint }$
such a unique state. 
\end{proposition}
 Proposition \ref{prop:uniquesol}
 is used in  \cite{de2024model}
 to prove the following theorem. 
\begin{theorem}[\cite{de2024model}]\label{theo:butcause}
    Let $\state = \tuple{\base,\val } \in \univCtxEq $  s.t. 
 its causal graph
$\causegraph_{\state }$
is a DAG,
$\term \in \Terms_{ \engfunct(\state ) }$
and $\Props(\omega )  \subseteq  \Props(\base ) $.
We have 
 $ 
  \butcause(\state,\term, \omega  )$
  iff
\begin{align*}
  & \exists \globint \in 
      \Interv{ \Props(\lambda ) }, \exists 
    \ArbProps \subseteq 
\engfunct(\state ),
\exists \globint'  \in  \Interv{\ArbProps } 
 \text{ such that } \\
&         \ArbProps \cap \Props(\lambda )=\emptyset,
\state \sat       \conjfunct{\globint' }
    \text{ and }
    \state^{\globint \cup \globint'  }
    \sat \neg \omega     . 
\end{align*}

\end{theorem}

\subsection{Proof of Lemma \ref{lem: key}}
\begin{proof}
We start from the right to left direction.
From the antecedent $\exists \lambda' \in \Terms_{\Props(\lambda)}$ s.t. $S \models 
(\lambda' \wedge \lambda^{\exgfunct}_{\state})
\diamondRight \neg \omega$, viz. $\exists S' \in \closest(\lambda' \wedge \lambda^{\exgfunct}_{\state}, S, \univCtx) $, $S' \models \neg \omega$.
To prove $S \models \butcause(S, \lambda, \phi)$, by Theorem \ref{theo:butcause} it is enough to construct some $E \in \Interv{\Props(\lambda)}, Z \subseteq end(S) \setminus \Props(\lambda), E' \in \Interv{\ArbProps} $ s.t. $S \models \widehat{E'}$ and $S^{E \cup E'} \models \neg \omega$.
Hence simply let $E$ be an event s.t. $\widehat{E} = \overline{\lambda}$, $E' = \{p \leftrightarrow \tau: p \in end(S) \setminus \Props(E), S \models p \leftrightarrow \tau, S' \models p \leftrightarrow \tau\}$, and $Z = \Props(E')$.
As $G_S$ is a DAG, by Proposition \ref{prop:uniquesol} $S^{E \cup E'}$ must exist.
We claim that $V'= V^{E \cup E'}$, therefore $S \models \widehat{E'}, S^{E \cup E'} \models \neg \omega$ (since values of propositional formulas are determined by valuation alone) and
$\butcause(S, \lambda, \omega)$ holds.

To prove the claim is to prove $V' \Delta V^{E \cup E'} = \emptyset$. $V' \Delta V^{E \cup E'} \cap \Props(E \cup E') = \emptyset$ by the construction of $E$ and $E'$. 
And $V' \Delta V^{E \cup E'} \cap \exgfunct(S) = \emptyset$, for $S' \models \lambda^{\exgfunct}_{\state}$, and $S^{E \cup E'}$ does not change the valuation on $\exgfunct(S)$.

Hence we know $V' \Delta V^{E \cup E'} \subseteq \big( \engfunct(S) \setminus \Props(E \cup E') \big)$. 
Consider the state $S'' = (C'', V'')$ defined as follows:
\begin{align*}
    V'' & = V^{E \cup E'}, \\
    C'' &= (C \cap (C' \cup C^{E \cup E'})) \setminus \{\ltri \omega \in C': V'' \not\models \omega \}.
\end{align*}
Clearly $S''$ is well defined, i.e. $V''$ is compatible with $C''$. We claim that $S'' \preceq_S S'$. For the valuation part, notice since $\big(V' \Delta V^{E \cup E'}\big) \cap \Props(E \cup E') = \emptyset$, and by definition $\Props(E') = \{p \in \engfunct(S) \setminus \Props(E): p \in V \iff p \in V'\}$, we have that if $p \in V' \Delta V^{E \cup E'}$ then $p \in V^{E\cup  E'} \iff p \in V$. Hence $V'' \Delta V \supseteq V' \Delta V $.
For the causal base part, let $D = (C \cap C') \setminus C''$. We aim to show that $D = \emptyset$. 
Suppose $D \neq \emptyset$ for a contradiction. Then since $\causegraph_{\state}$ is a DAG and $V'' \cap \exgfunct(S) = V' \cap \exgfunct(S)$, there must be a $p' \leftrightarrow \omega' \in D$ with $\Props(\omega') \cap (V'' \Delta V') = \emptyset$, otherwise $\causegraph_{\state}$ would be defeated by a loop.
That means $V'' \models \omega' \iff V' \models \omega'$. W.l.o.g., assume $V'' \models \omega'$ and $V' \models \omega'$. Since $p' \leftrightarrow \omega' \in C'$, we have $V' \models p'$. 
Now note that $p' \in \Props(E \cup E')$, for $C^{E \cup E'}$ only removes a rule if the rule-head is in $E \cup E'$. 
By construction of $E, E'$ we shall have $p \in V' \iff p \in V''$, therefore $V'' \models p'$. That means $V'' \models p' \leftrightarrow \omega'$, i.e. the rule is compatible with $V''$ and $p' \leftrightarrow \omega' \in C''$, contradicting that $p' \leftrightarrow \omega' \in D$.
Therefore if $V' \neq V^{E \cup E'}$, we would have $S'' \prec_S S'$, and $S'' \models \lambda' \wedge \lambda^{\exgfunct}_{\state}$, contradicting $S' \in \closest(\lambda' \wedge \lambda^{\exgfunct}_{\state})$ our assumption. Thus the claim $V = V^{E \cup E'}$ is proven.

For the other direction suppose $S \models \lambda$ and $\butcause(S, \lambda, \phi)$. Then there are $E, E'$ as in Definition \ref{def:Butcause} s.t. $S \models \widehat{E'}$ and, since $\causegraph_{\state}$ is a DAG, by Proposition \ref{prop:uniquesol} and Theorem \ref{theo:butcause}, $S^{E \cup E'}$ exists and $S^{E \cup E'} \models \lambda^{\exgfunct}_{\state} \wedge \neg \omega$.
We let $\lambda' = \widehat{E}$.
To prove $S \models (\lambda' \wedge \lambda^{\exgfunct}_{\state}) \diamondRight \neg \omega$, we claim $\exists S' \in \closest( \lambda' \wedge \lambda^{\exgfunct}_{\state}, S, \univCtx)$ s.t. $V' = V^{E \cup E'}$. If so, then $S' \models \neg \omega$ as we want.

Suppose the claim not holds towards a contradiction, then $\forall S^* \in \univCtx$ with $S^* \preceq_S S'$ and $V^* = V^{E \cup E'}$, $\exists S' \in \closest( \lambda' \wedge \lambda^{\exgfunct}_{\state} , S, \univCtx), S' \prec_S S^* \preceq_S S^{E \cup E'}$ and $V^{E \cup E'} \Delta V \supset V' \Delta V $. Let $D = (V^{E \cup E'} \Delta V) \setminus  (V' \Delta V) \neq \emptyset$.
We observe $\forall p_i \in  D$, $p_i \in end(S) \setminus \Props(E \cup E')$. 
To see that, clearly $p_i \notin exo(S)$ by construction of $V^{E \cup E'}$; $p_i \notin \Props(E)$ for both $S^{E \cup E'}$ and $S'$ satisfy $\widehat{E}$; $p_i \notin \Props(E')$ for both $S, S^{E \cup E'}$ satisfy $\widehat{E'}$. 
Then $\forall p_i \in D$, there must be some $p_i \leftrightarrow \omega_i \in C \cap C^{E \cup E'}$, because $C^{E \cup E'}$ only removes the rules for variables in $E \cup E'$.
Hence also $p_i \leftrightarrow \omega_i \in C'$, otherwise $C^{E \cup E'} \cap C \not\subseteq C' \cap C$, which fails $S' \prec_S S^{E \cup E'}$.
     Again since the graph of $S$ is a DAG, we could assume some $p_i \leftrightarrow \omega_i \in C \cap C^{E \cup E'} \cap C'$ s.t. $D\cap \Props(\omega_i) = \emptyset$. Otherwise the DAG will be defeated by a loop. 
But that gives $S' \models \omega_i$ iff $S^{E \cup E'} \models \omega_i$, forcing $p_i \in V'$ iff $p_i \in V^{E \cup E'}$, a contradiction wanted.
\end{proof}

\subsection{Proof of Theorem \ref{thm:actual_cause_minimal_might} }
\begin{proof}
    For the left to right direction, let $\lambda$ be an actual cause. Therefore $\butcause(S, \lambda, \omega)$ holds, by Lemma \ref{lem: key} $S \models \bigvee_{\lambda' \in \Terms_{\Props(\lambda)} } ((\lambda' \wedge \lambda^{\exgfunct}_{\state}) \diamondRight \neg \omega)$.
Suppose for contradiction $\exists \lambda' \in \Terms_{\Props(\lambda)}$ s.t. $(\lambda' \wedge \lambda^{\exgfunct}_{\state}) \diamondRight \neg \omega$ and $\lambda' \neq \overline{\lambda}$. If $|\lambda'| < |\lambda|$, which by Lemma \ref{lem: key} means $\butcause(S, \lambda_1, \omega)$ holds for some $\lambda_1 \subset \lambda$ with $\Props(\lambda_1) = \Props(\lambda')$, then $\lambda$ is not a minimal ``but'' condition, a contradiction.
If $|\lambda'| = |\lambda|$, then $\exists \ell \subseteq (\lambda' \cap \lambda) \setminus \overline{\lambda} $, where $\ell$ denotes a literal. Let $\lambda'' = \lambda' \setminus \ell$. Again since $\lambda$ is a minimal ``but'' condition for $\omega$ at $S$, by Lemma \ref{lem: key} we have $S \models (\lambda'' \wedge \lambda^{\exgfunct}_{\state}) \boxRight \omega $. Now by Validity (\ref{validity:termRM1}) or (\ref{validity:termRM2}) in Proposition \ref{prop: validities} (depending on whether $\ell$ is positive or negative), together with $S \models \ell$ we must have $S \models (\ell \wedge \lambda'' \wedge \lambda^{\exgfunct}_{\state}) \boxRight \omega $. But $\ell \wedge \lambda''$ is nothing but $\lambda'$, hence $S \models (\lambda' \wedge \lambda^{\exgfunct}_{\state}) \boxRight \omega$, a contradiction as well. So, it has to be $S \models (\overline{\lambda} \wedge \lambda^{\exgfunct}_{\state}) \diamondRight \neg \omega$. 
    To show 
    $S \models \bigwedge_{\lambda' \in \Terms_{\ArbProps}, \ArbProps \subset \Props(\lambda)} (\lambda' \boxRight \omega)$
    suppose for contradiction $S \models (\lambda' \wedge \lambda^{\exgfunct}_{\state}) \diamondRight \neg \omega$ for a $\lambda'$ with $\Props(\lambda') \subset \Props(\lambda)$. Then
    $\exists \lambda_1 \subset \lambda$, 
    s.t. $\Props(\lambda_1) = \Props(\lambda')$.
    By Lemma \ref{lem: key} $\butcause(S, \lambda_1, \omega)$ holds, contradicting the minimality of $\lambda$.  

    Now assume the right hand side holds, we prove $\lambda$ is an actual cause.
    By Lemma \ref{lem: key} from 
    $S \models \lambda \wedge (\overline{\lambda} \wedge \lambda^{\exgfunct}_{\state})  \diamondRight \neg \omega$
    we obtain $\butcause(S, \lambda, \omega)$. To show that $\lambda$ is subset-minimal, suppose towards a contradiction $\exists \lambda_1 \subset \lambda$, that $\butcause(S, \lambda_1, \omega)$ holds. Then by Lemma \ref{lem: key} we have 
    $S \models (\lambda' \wedge \lambda^{\exgfunct}_{\state}) \diamondRight \neg \omega $ for some $\lambda' \in \Terms_{\Props(\lambda_1)}$,
    contradicting the assumption.
\end{proof}

\section{Model Checking}

\label{sec:QBF_encodings}
\subsection{Definition of QBF Satisfiability}

The language of QBF formulas is given by the following grammar.
\[ \langQBF \quad\isdef\quad \tau \coloncolonequals \prop \in \Props \mid \top \mid \lnot\tau
	\mid \tau\land\tau
	\mid \forallqbf{\prop}{\tau} ,
	\\ \]

Operators  $\bot$, $\lor$, $\limp$, $\lequ$ and $\exists$ are defined as usual abbreviations. A formula is called \emph{closed} if every variable in it is bound by some quantifier, and \emph{open} otherwise.

Satisfaction of QBF formulas on valuations $\val \subseteq \Props$ is defined as follows:
\[\begin{array}{lll}
  \val \modelsQBF \prop   
  & \textit{ iff } &\prop \in  \val,
  \\
  \val  \modelsQBF \top   
  & \textit{ is true, }
  \\
  \val  \modelsQBF \neg \tau   
  & \textit{ iff } & \val \not\modelsQBF \tau,
  \\
  \val \modelsQBF \tau_1 \land \tau_2   
  & \textit{ iff } & \val \modelsQBF \tau_1 \textit{ and } \\
  & & \val \modelsQBF \tau_2,
  \\
  \val  \modelsQBF  \forallqbf{\prop}{\tau}   
  & \textit{ iff } & (\val \setminus \{\prop\}) \modelsQBF \tau \textit{ and } \\
  && (\val \cup \{\prop\}) \modelsQBF \tau.
  \\
\end{array}\]

Note that satisfaction of a closed formula $\tau$ does not depend on the valuation, in such case it is said that $\tau$ is true (denoted $\modelsQBF \tau$). 

\begin{framed}
\noindent\textbf{QBF problem.} \\
Input: closed $\tau \in \langQBF$. \\
Output: \texttt{true} if $\modelsQBF \tau$, \texttt{false} otherwise.
\end{framed}

\subsection{QBF Encoding of Modal Checking}
\label{sec:QBF_encodings_modal_checking}

The main predicate for model checking:

\[
\begin{array}{lcl}
\qbfSat(\prop, \stateVarsi) & = &
\statevarvali \\
\qbfSat(\Delta \omega, \stateVarsi) & = & \statevarbasei  \\
\qbfSat(\neg \varphi, \stateVarsi) & = & \neg \qbfSat(\varphi, \stateVarsi) \\
\qbfSat(\varphi_1 \land \varphi_1, \stateVarsi) & = & \qbfSat(\varphi_1, \stateVarsi) \land \qbfSat(\varphi_2, \stateVarsi) \\
\qbfSat(\condmodal{\varphi_1}{\varphi_2}, \stateVarsi) & = & \forall \stateVars{\qbfindi+1}. \; \qbfState(\stateVars{\qbfindi+1}) \rightarrow \\
 & &  \qquad\qbfClosest(\varphi_1, \stateVarsi, \stateVars{i+1}) \rightarrow \\
 & & \qquad\qbfSat(\varphi_2, \stateVars{\qbfindi+1}) \\
\end{array}
\]

\phantom{   } 

\phantom{   } 

Auxiliary predicates:
\[
\begin{array}{l}
\qbfClosest(\varphi, \stateVarsi, \stateVarsii) = \forall \stateVarsiii.\; \forall r.\; \qbfState(\stateVarsiii) \rightarrow \\
\hspace{37mm} (\qbfSat(\varphi, \stateVarsiii) \leftrightarrow r) \rightarrow  \\
\hspace{37mm}((\qbfEq(\stateVarsii, \stateVarsiii) \rightarrow r) \wedge \\
\hspace{37mm} (\qbfCloser(\stateVarsi, \stateVarsii, \stateVarsiii) \rightarrow
\neg r)).  
\end{array} \]
\noindent where $k = \max(\{i,j\})+1$.
\[ \begin{array}{lcl}
\qbfCloser(\stateVarsi, \stateVarsii, \stateVarsiii) & = & \qbfClosereq(\stateVarsi, \stateVarsii, \stateVarsiii) \wedge \\
& & (\neg \qbfClosereq(\stateVarsi, \stateVarsiii, \stateVarsii)) \\
\end{array} \]
\[ \begin{array}{lcl}
\qbfClosereq(\stateVarsi, \stateVarsii, \stateVarsiii) & = & \bigwedge\limits_{\omega \in \Gamma} ((\statevarbasei \land \statevarbaseii) \rightarrow \statevarbaseiii) \wedge \\
& & \bigwedge\limits_{p \in \Sigma} (((\statevarvali \land \neg \statevarvaliii) \rightarrow \neg \statevarvalii) \wedge \\
& & \phantom{\bigwedge\limits_{p \in \Sigma} (} ((\neg \statevarvali \land \statevarvaliii) \rightarrow \statevarvalii))\\
\end{array} \]
\[ \begin{array}{lcl}
\qbfEq(\stateVarsii, \stateVarsiii) & = & \bigwedge\limits_{\omega \in \Gamma} (\statevarbaseii \leftrightarrow \statevarbaseiii) \wedge 
\bigwedge\limits_{p \in \Sigma} (\statevarvalii \leftrightarrow \statevarvaliii) \\
\end{array} \]
\[ \begin{array}{lcl}
\qbfState(\stateVarsi) & = & \bigwedge_{\omega \in \Gamma} (\statevarbasei \rightarrow \qbfSat(\omega, \stateVarsi))
\end{array}
\]

\subsection{Proof of Theorem \ref{th:model_checking_lower_bound} }

\begin{proof} 

\textbf{PSPACE-membership}:

The encoding of model checking into QBF is given by the following fact:
$\tuple{\state, \univCtxRel{\Gamma}} \models \inputform$
iff a closed QBF formula \[ \tau_0 =  \exists X^0. \qbfInit(\stateVars{0}, \state) \wedge \qbfSat(\inputform, \stateVars{0})
 \] is true, where 
\begin{align*}
\qbfInit(\stateVars{0}, \tuple{\base,\val}) = \bigwedge\limits_{\omega\in\base} \statevarbase{0}{\omega} \wedge \bigwedge\limits_{\omega\notin\base} \neg\statevarbase{0}{\omega} \wedge \\
\bigwedge\limits_{\prop\in\val\cap\Sigma} \statevarval{0}{\prop} \wedge \bigwedge\limits_{\prop\notin\val\cap\Sigma} \neg\statevarval{0}{\prop}.
\end{align*}

To prove it we need to check correctness of all predicates we define (see Section~\ref{sec:QBF_encodings_modal_checking}).

We recall the following notion from Section~\ref{sec:mc}, which reconstructs a state (or, in the general case, tuple of causal base and valuation that are not necessarily consistent) from their encodings by variables $\stateVarsi$: for $\qbfVal \subseteq \stateVarsi$,  $\qbfValToState{\qbfindi}{\qbfVal} = \tuple{\{ \omega \mid \statevarbasei \in \qbfVal \}, \{ \prop \mid \statevarvali \in \qbfVal \}}$. We will also use notation $\subformulas{\inputform}$ to refer to the set of subformulas of the input formula $\inputform$.

We formulate and prove the correctness of predicates one by one.
\begin{enumerate}
\item\label{item:qbf_corr_propsat} For any $\omega \in \Gamma$, and any $\qbfVali \subseteq \stateVarsi$, if $\qbfValToState{\qbfindi}{\qbfVali} = \tuple{\base, \val}$ then $\qbfVali \modelsQBF \qbfSat(\omega, \stateVarsi)$ iff $\val \models \omega$. It can be proved by trivial induction on $\omega$.
\item\label{item:qbf_corr_state} For any $\qbfVali \subseteq \stateVarsi$, $\qbfVali \modelsQBF \qbfState(\stateVarsi)$ iff $\qbfValToState{\qbfindi}{\qbfVali} \in \univCtxRel{\Gamma}$. Using the fact (\ref{item:qbf_corr_propsat}) above, $\qbfVali \modelsQBF \qbfState(\stateVarsi)$ is equivalent to the fact that $\val \models \omega$ (where $\val$ is the valuation of $\qbfValToState{\qbfindi}{\qbfVali}$) for any $\omega \in \Gamma$ such that $\statevarbasei \in \qbfVali$. This is exactly the consistency condition for the state $\qbfValToState{\qbfindi}{\qbfVali}$.
\item\label{item:qbf_corr_eq} For any $\qbfValii \subseteq \stateVarsii$, $\qbfValiii \subseteq \stateVarsiii$, $\qbfValii \cup \qbfValiii \modelsQBF \qbfEq(\stateVarsii, \stateVarsiii)$ iff $\qbfValToState{\qbfindii}{\qbfValii} = \qbfValToState{\qbfindiii}{\qbfValiii}$. This holds trivially.
\item\label{item:qbf_corr_closereq} For ${\qbfVali \subseteq \stateVarsi}$, ${\qbfValii \subseteq \stateVarsii}$, ${\qbfValiii \subseteq \stateVarsiii}$, ${\qbfVali \cup \qbfValii \cup \qbfValiii \modelsQBF \qbfClosereq(\stateVarsi, \stateVarsii, \stateVarsiii)}$ iff ${\qbfValToState{\qbfindii}{\qbfValii} \preceq_{\qbfValToState{\qbfindi}{\qbfVali}} \qbfValToState{\qbfindiii}{\qbfValiii}}$.  The first conjunct in $\qbfClosereq(\stateVarsi, \stateVarsii, \stateVarsiii)$ encodes the fact $\base^{\qbfindi} \cap \base^{\qbfindii} \subseteq \base^{\qbfindiii}$ and the second conjunct encodes the fact $\val^{\qbfindi} \Delta \val^{\qbfindiii} \subseteq \val^{\qbfindi} \Delta \val^{\qbfindii}$ (for ${\qbfValToState{\qbfindi}{\qbfVali} = \tuple{\base^{\qbfindi},\val^{\qbfindi}}}$, ${\qbfValToState{\qbfindii}{\qbfValii} = \tuple{\base^{\qbfindii},\val^{\qbfindii}}}$, ${\qbfValToState{\qbfindiii}{\qbfValiii} = \tuple{\base^{\qbfindiii},\val^{\qbfindiii}}}$), which is exactly the Definition~\ref{def:similarity_relation} of similarity relation.
\item\label{item:qbf_corr_closer} For ${\qbfVali \subseteq \stateVarsi}$, ${\qbfValii \subseteq \stateVarsii}$, ${\qbfValiii \subseteq \stateVarsiii}$, ${\qbfVali \cup \qbfValii \cup \qbfValiii \modelsQBF \qbfCloser(\stateVarsi, \stateVarsii, \stateVarsiii)}$ iff ${\qbfValToState{\qbfindii}{\qbfValii} \prec_{\qbfValToState{\qbfindi}{\qbfVali}} \qbfValToState{\qbfindiii}{\qbfValiii}}$. Trivially by definition of $\prec_{\qbfValToState{\qbfindi}{\qbfVali}}$ (Definition~\ref{def:sat}), relying on the fact~\ref{item:qbf_corr_closereq} above.
\item\label{item:qbf_corr_sat_closest} The correctness of $\qbfClosest(\varphi, \stateVarsi, \stateVarsii)$ and $\qbfSat(\varphi, \stateVarsi)$ we need to prove together by mutual induction.
\renewcommand{\labelenumii}{\arabic{enumi}.\arabic{enumii}.}
\begin{enumerate}
\item\label{item:qbf_corr_sat} For $\varphi \in \subformulas{\inputform}$, ${\qbfVali \subseteq \stateVarsi}$,  ${\qbfVali \modelsQBF \qbfSat(\varphi, \stateVarsi)}$ iff ${\qbfValToState{\qbfindi}{\qbfVali} \models \varphi}$.
\item\label{item:qbf_corr_closest} For $\varphi \in \subformulas{\inputform}$, ${\qbfVali \subseteq \stateVarsi}$, ${\qbfValii \subseteq \stateVarsii}$, ${\qbfVali \cup \qbfValii \modelsQBF \qbfClosest(\varphi, \stateVarsi, \stateVarsii)}$ iff $\qbfValToState{\qbfindii}{\qbfValii} \in \closest(\varphi, \qbfValToState{\qbfindi}{\qbfVali}, \univCtxRel{\Gamma})$.
\end{enumerate}

The fact (\ref{item:qbf_corr_sat}) follows directly by definition from the inductive hypotheses about correctness of $\qbfSat$ for immediate subformulas and about correctness of $\qbfClosest$ for the antecedent in the case of counterfactual.

For the fact (\ref{item:qbf_corr_closest}) notice that the definition of $\closest$ prohibits for each valuation on $\stateVarsiii$ the situation when $\qbfState(\stateVarsiii)$ is true, $\qbfEq(\stateVarsii, \stateVarsiii)$ is true and $\qbfSat(\varphi, \stateVarsiii)$ is false and the situation when $\qbfState(\stateVarsiii)$ and $\qbfCloser(\stateVarsi, \stateVarsii, \stateVarsiii)$ and $\qbfSat(\varphi, \stateVarsiii)$ are all true, and no other situations. Using facts (\ref{item:qbf_corr_state}), (\ref{item:qbf_corr_eq}), (\ref{item:qbf_corr_closer}) above, and the fact (\ref{item:qbf_corr_sat}) for $\varphi$, we get from that exactly ${\qbfValToState{\qbfindii}{\qbfValii} \models \varphi}$ (since the first situation is prohibited) and the fact that $\qbfValToState{\qbfindiii}{\qbfValiii} \in \univCtxRel{\Gamma}$ and ${\qbfValToState{\qbfindii}{\qbfValii} \prec_{\qbfValToState{\qbfindi}{\qbfVali}} \qbfValToState{\qbfindiii}{\qbfValiii}}$ imply $\qbfValToState{\qbfindiii}{\qbfValiii} \not\models \varphi$ (since the second situation is prohibited) with $\qbfValToState{\qbfindiii}{\qbfValiii}$ going through all possible states. This is exactly the definition of the closest state.
\end{enumerate}

Now we are in a position to prove that $\tau_0$ is satisfiable iff $\tuple{\state, \univCtxRel{\Gamma}} \models \inputform$. $\qbfInit(\stateVars{0}, (\base,\val))$ is true exactly on valuation $Y^0$ for which $\qbfValToState{0}{Y^0} = (\base, \val)$. So satifsability of $\tau_0$ is equivalent to $\tuple{\state, \univCtxRel{\Gamma}} \models \inputform$ by the fact (\ref{item:qbf_corr_sat}) for $\inputform$.

It only remains to check that encoding has polynomial size w.r.t. the input. We have the following estimations for the sizes of the predicates (the second and the last case by trivial structural induction on formula):
\begin{itemize}
\item $|\qbfInit(\stateVarsi, S)| = \O{|S|}$
\item $|\qbfSat(\omega, \stateVarsi)| = \O{|\omega|}$ for $\omega \in \Gamma$
\item $|\qbfState(\stateVarsi)| = \O{\sum_{\omega \in \Gamma} |\omega|}$
\item $|\qbfEq(\stateVarsii, \stateVarsiii)| = \O{|\Gamma| + |\Sigma|}$
\item $|\qbfClosereq(\stateVarsi,\stateVarsii, \stateVarsiii)| = \O{|\Gamma| + |\Sigma|}$
\item $|\qbfCloser(\stateVarsi,\stateVarsii, \stateVarsiii)| = \O{|\Gamma| + |\Sigma|}$
\item $|\qbfSat(\varphi,\stateVarsi)| = \\ |\qbfClosest(\varphi,\stateVarsi,\stateVarsii)| = {\O{|\varphi| \cdot (|\Sigma| + \sum_{\omega \in \Gamma} |\omega|)}}$
\end{itemize}

So the reduction output $\tau_0$ indeed has a polynomial size w.r.t. the input.

\textbf{PSPACE-hardness}: We provide a reverse reduction. Suppose that we are given QBF formula $\tau_0$.
We assume that $\tau_0$ is closed (every variable occurring in $\tau$ is bounded by a quantifier) and that different quantifiers in it quantify over different variables from $\Props$ (this can be easily achieved by renaming bounded variables).  For the encoding we will pair each of the bounded variable $x$ in $\tau_0$ by a ``fresh'' variable $\cloneVar(x)$, such that all these variables are different (i.e. $\cloneVar$ is injective) and do not occur in $\tau_0$ (such $\cloneVar$ can be easily chosen since $\Props$ is infinite). Then we have the following encoding for the subformulas of $\tau_0$:
\[ \begin{array}{lcl}
t(\prop) & = & \prop \\
t(\lnot \tau) & = & \lnot t(\tau) \\
t(\tau_1 \land \tau_2) & = & t(\tau_1) \land t(\tau_2) \\
t(\forallqbf{\prop}{\tau}) & = & \condmodal{(\prop \lor \cloneVar(\prop))}{t(\tau)} \\
\end{array} \]
We can prove by structural induction that for every subformula $\tau$ of $\tau_0$ if valuation $V$ does not contain neither $\prop$ nor $\cloneVar(\prop)$ for any variable $\prop$ bounded in $\tau$ then $V \modelsQBF \tau$ iff $((\emptyset, \val), \univCtxRel{\emptyset}) \models t(\tau)$. Base case is trivial, cases of propositional connectives follow directly from the induction hypotheses. If $\tau = \forallqbf{\prop}{\tau'}$ we have $\{\prop, \cloneVar(\prop)\} \cap \val = \emptyset$ since $\prop$ is bounded in $\tau$. ${\closest(\prop \lor \cloneVar(\prop),(\emptyset,\val),\univCtx)} = \{ \tuple{\emptyset, \val \cup \{\prop\}}, (
\tuple{\emptyset, \val \cup \{\cloneVar(\prop)\}} \}$, since $(\prop \lor \cloneVar(\prop))$ is not satisfied in $(\emptyset,\val)$ but is satisfied in these two states, and for every other state in $\univCtxRel{\emptyset}$ satisfying $(\prop \lor \cloneVar(\prop))$ the valuation contains either $\prop$ or $\cloneVar(\prop)$, so one of the states above is strictly more similar to $(\emptyset, \val)$. Therefore $((\emptyset, V), \univCtx) \models t(\forallqbf{\prop}{\tau'})$ iff both ${((\emptyset, V \cup \{\prop\}),\univCtx) \models t(\tau')}$ and ${((\emptyset, V \cup \{\cloneVar(\prop)\}),\univCtx) \models t(\tau')}$; by the inductive hypotheses the first is equivalent to ${(V \cup \{ \prop \}) \modelsQBF \tau'}$ and second is equivalent to $(V \cup \{ \cloneVar(\prop) \}) \modelsQBF \tau'$ which is also equivalent to $V \modelsQBF \tau'$ (since $\cloneVar(\prop)$ does not occur in $\tau'$), and their conjunction is equivalent exactly to $V \modelsQBF \forallqbf{\prop}{\tau'}$ (since $\prop \not\in V$ by the initial assumption on $V$). 

As a result, closed QBF formula $\tau_0$ evaluates to true iff $((\emptyset, \emptyset), \univCtx) \models t(\tau_0)$. Since $|t(\tau_0)| = \mathcal{O}(|\tau_0|)$, the encoding implies PSPACE-hardness of model checking. 
\end{proof}

\subsection{Proof of Lemma \ref{lem:mc_relativized_reduction_unnested} }

\begin{proof}
We first generalize the statement: instead of fixing $\Gamma$, we prove that $\tuple{\tuple{\base, \val}, \univCtx} \models \inputform$ is equivalent to $\tuple{\tuple{\base, \val}, \univCtxRel{\Gamma}} \models \inputform$ for any $\Gamma$ such that $\base \subseteq \Gamma$ and  $\omega \in \Gamma$ for any subfomula $\ltri \omega$ in $\inputform$.
We prove this generalized statement by structural induction on formula $\inputform$.

\begin{itemize}
\item $\inputform = p \in \Props$: Trivially.
\item $\inputform = \top$: Trivially.
\item $\inputform = \Delta \omega$: By the assumption for $\Gamma$.
\item $\inputform = \neg \varphi_1$: By the inductive hypothesis.
\item $\inputform = \varphi_1 \land \varphi_2$: By the inductive hypotheses.
\item $\inputform = \condmodal{\alpha}{\beta}$: Since $\inputform$ does not contain nested counterfactual by assumption, $\alpha, \beta \in \langE$. 

We will use the surjective mapping $\maprel \colon \univCtx \rightarrow \univCtxRel{\Gamma}$ defined as $\maprel(\tuple{\base^\ast, \val^\ast}) = \tuple{\base^\ast \cap \Gamma, \val^\ast}$. We will need the following fact about $\maprel$: for $\gamma \in \{\alpha, \beta\}$, for any $\state \in \univCtx$,
$\tuple{\maprel(\state), \univCtxRel{\Gamma}} \models \gamma$ iff $\tuple{\state, \univCtx} \models \gamma$. It is true since evaluation on these models is the same for all variables (since the valuation is the same) and for all subformulas of the form $\ltri \omega$ in $\gamma$ (since $\omega \in \Gamma$ by the assumption on $\Gamma$), and since $\gamma \in \langE$ the models do not change in the evaluation, so it is sufficient for equivalent evaluation of $\gamma$ on these two models. Another fact about $\maprel$ we will need: for any $\state_1, \state_2 \in \univCtx$, ${\state_1 \prec_{\tuple{\base, \val}} \state_2}$ iff ${\maprel(\state_1) \prec_{\tuple{\base, \val}} \maprel(\state_2)}$. It is true due to the assumption $C \subseteq \Gamma$.

Let us now prove the inductive case. First, let us unfold the definition of ${\tuple{\tuple{\base, \val}, \univCtx} \models \condmodal{\alpha}{\beta}}$:
\begin{multline*} \forall \state_1 \in \univCtx.\; \tuple{\state_1, \univCtx} \models \beta \;\vee\; \tuple{\state_1, \univCtx} \not\models \alpha \;\vee\; \\
(\exists \state_2 \in \univCtx.\; \state_1 \!\prec_{\tuple{\base, \val}}\! \state_2 \;\wedge\; \tuple{\state_2, \univCtx} \models \alpha). \end{multline*}

Using the facts about $\maprel$ we can rewrite it as follows:
\begin{multline*} \forall \state_1 \in \univCtx.\; \tuple{\maprel(\state_1), \univCtxRel{\Gamma}} \models \beta \;\vee\; \tuple{\maprel(\state_1), \univCtxRel{\Gamma}} \not\models \alpha \;\vee\; \\
(\exists \state_2 \in \univCtx.\; \maprel(\state_1) \!\prec_{\tuple{\base, \val}}\! \maprel(\state_2) \;\wedge\; \tuple{\maprel(\state_1), \univCtxRel{\Gamma}} \models \alpha). \end{multline*}

Now, since $\maprel$ is a surjective mapping onto $\univCtxRel{\Gamma}$, quantifying over $\state \in \univCtx$ and using only $\maprel(\state)$ is equivalent to quantifying over $\state' \in \univCtxRel{\Gamma}$ and using $\state'$ directly. We can perform this rewriting for both quantifiers:
\begin{multline*} \forall \state'_1 \in \univCtxRel{\Gamma}.\; \tuple{\state'_1, \univCtxRel{\Gamma}} \models \beta \;\vee\; \tuple{\state'_1, \univCtxRel{\Gamma}} \not\models \alpha \;\vee\; \\
(\exists \state'_2 \in \univCtxRel{\Gamma}.\; \state'_1 \!\prec_{\tuple{\base, \val}}\! \state'_2 \;\wedge\; \tuple{\state'_2, \univCtxRel{\Gamma}} \models \alpha). \end{multline*}

Which is equivalent to ${\tuple{\tuple{\base, \val}, \univCtxRel{\Gamma}} \models \condmodal{\alpha}{\beta}}$.\qedhere
\end{itemize}
\end{proof}

\subsection{QBF Encoding of Actual Cause}

Similarly to the encodings of states, we introduce sets of variables $\litVars{i}$ to encode terms: for $i \in \Nats$, $\litVars{i} = {\{ \litvarpos{i}{\prop} \mid \prop \in \Sigma \}} \cup {\{ \litvarneg{i}{\prop} \mid \prop \in \Sigma \}}$. To encode a term the variables should not have contradicting literals:
\[ \qbfTerm(\litVars{i}) = \bigwedge_{\prop \in \Sigma} \neg (\litvarpos{i}{\prop} \wedge \litvarneg{i}{\prop}))  \]

We first define some operations on terms:

\[\qbfTermRev(\litVars{i}, \litVars{j}) = \bigwedge_{\prop \in \Sigma} ((\litvarpos{i}{\prop} \leftrightarrow \litvarneg{j}{\prop}) \wedge (\litvarneg{i}{\prop} \leftrightarrow \litvarpos{j}{\prop}))\]

\[\begin{array}{l}
\qbfTermMerge(\litVars{i}, \litVars{j}, \litVars{k}) = \bigwedge\limits_{\prop \in \Sigma} ((\litvarpos{i}{\prop} \vee \litvarpos{j}{\prop} \rightarrow \litvarpos{k}{\prop}) \wedge \\ \phantom{\qbfTermMerge(\litVars{i}, \litVars{j}, \litVars{k}) = \bigwedge_{\prop \in \Sigma}} (\litvarneg{i}{\prop} \vee \litvarneg{j}{\prop} \rightarrow \litvarneg{k}{\prop})) 
\end{array} \]

\[\begin{array}{l}
\qbfTermVarsSubseteq(\litVars{i}, \litVars{j}) = \bigwedge\limits_{\prop \in \Sigma} ((\litvarpos{i}{\prop} \vee \litvarneg{i}{\prop}) \rightarrow (\litvarpos{j}{\prop} \vee \litvarneg{j}{\prop})) 
\end{array} \]

\[\begin{array}{l}
\qbfTermVarsSubset(\litVars{i}, \litVars{j}) = \qbfTermVarsSubseteq(\litVars{i}, \litVars{j}) \wedge \\
\phantom{\qbfTermVarsSubset(\litVars{i}, \litVars{j}) =} \neg \qbfTermVarsSubseteq(\litVars{j}, \litVars{i}) 
\end{array} \]

Now auxiliary predicates involving states:

\[\begin{array}{l}
\qbfTermExo(\litVars{i}, \stateVars{j}) = \bigwedge\limits_{\prop \in \exgfunct(\state )} ((\litvarpos{i}{\prop} \leftrightarrow \statevarval{j}{\prop}) \wedge (\litvarneg{i}{\prop} \leftrightarrow \neg \statevarval{j}{\prop})) \\
\phantom{\qbfTermExo(\litVars{i}, \stateVars{j})} \wedge 
 \bigwedge\limits_{\prop \in \Sigma \setminus \exgfunct(\state )} (\neg \litvarpos{i}{\prop} \wedge \neg \litvarneg{i}{\prop})
\end{array}\]

\[\qbfTermSat(\litVars{i}, \stateVars{j}) = \bigwedge_{\prop \in \Sigma} ((\litvarpos{i}{\prop} \rightarrow \statevarval{j}{\prop}) \wedge (\litvarneg{i}{\prop} \rightarrow \neg \statevarval{j}{\prop}))\]

\[
\begin{array}{l}
\qbfTermClosest(\litVars{t}, \stateVars{i}, \stateVars{j}) = \forall \stateVars{k}.\; \forall r.\; \qbfState(\stateVars{k}) \rightarrow \\
\hspace{37mm} (\qbfTermSat(\litVars{t}, \stateVars{k}) \leftrightarrow r) \rightarrow  \\
\hspace{37mm}((\qbfEq(\stateVarsii, \stateVarsiii) \rightarrow r) \wedge \\
\hspace{37mm} (\qbfCloser(\stateVarsi, \stateVarsii, \stateVarsiii) \rightarrow
\neg r)), 
\end{array} \] \noindent where $k = \max(\{t,i,j\})+1$.

\[
\begin{array}{l}
\qbfTermCounterfactual(\litVars{t}, \omega, \stateVarsi) = \forall \stateVars{n}. \; \qbfState(\stateVars{n}) \rightarrow \\
 \phantom{\forall \stateVars{\qbfindi+1}.\;} \qbfTermClosest(\litVars{t}, \stateVarsi, \stateVars{n}) \rightarrow \\
 \phantom{\forall \stateVars{\qbfindi+1}.\;} \qbfSat(\omega, \stateVars{n}), \\
\end{array}
\]
\noindent where $n = \max(\{t,i\})+1$.

Now we can define the main predicate verifying actual cause according to Theorem~\ref{thm:actual_cause_minimal_might}.

\[
\begin{array}{l}
\qbfActualCause(\litVars{t}, \omega, \stateVars{i}) = \qbfTermSat(\litVars{t}, \stateVars{i}) \wedge \\
\quad (\exists \litVars{n}.\; \exists \litVars{n+1}.\; \exists \litVars{n+2}.\; \\
\qquad \qbfTermRev(\litVars{n}, \litVars{t}) \rightarrow \\
\qquad \qbfTermExo(\litVars{n+1}, \stateVars{i}) \rightarrow \\
\qquad \qbfTermMerge(\litVars{n}, \litVars{n+1},\litVars{n+2}) \rightarrow \\
\qquad  \neg\qbfTermCounterfactual(\litVars{n+2}, \omega, \stateVars{i}) ) \\
\quad (\forall \litVars{n}.\; \forall \litVars{n+1}.\; \forall \litVars{n+2}.\; \\
\qquad \qbfTerm(\litVars{n}) \rightarrow \\
\qquad \qbfTermVarsSubset(\litVars{n}, \litVars{t}) \rightarrow \\
\qquad \qbfTermExo(\litVars{n+1}, \stateVars{i}) \rightarrow \\
\qquad \qbfTermMerge(\litVars{n}, \litVars{n+1},\litVars{n+2}) \rightarrow \\
\qquad  \qbfTermCounterfactual(\litVars{n+2}, \omega, \stateVars{i}) ),
\end{array}
\]

\noindent where $n = \max(\{t,i\})+1$.

\end{document}